\documentclass[12pt,preprint]{aastex}
 
\newcommand{\sdss}{{\small {SDSS}}}
\newcommand{\sdssi}{{\small {SDSS-I}}}
\newcommand{\sdssii}{{\small {SDSS-II}}}

\newcommand{\snls}{{\small {SNLS}}}
\newcommand{\essence}{{\small {ESSENCE}}}
\newcommand{\sn}{{\small {SN}}}
\newcommand{\snia}{{\small {SN~I}}a}
\newcommand{\snibc}{{\small {SN~I}}b/c}
\newcommand{\snii}{{\small {SN~II}}}
\newcommand{\sne}{{\small {SN}}e}
\newcommand{\sneia}{{\small {SN}}e{\small{~I}}a}

\newcommand{\photo}{{\tt {Photo}}}
\newcommand{\suspect}{{\small {SUSPECT}}}
\newcommand{\dmB}{$\Delta m_{15}(B)$}

\newcommand{\ccd}{{\small {CCD}}}
\newcommand{\apo}{{\small {APO}}}
\newcommand{\agn}{{\small {AGN}}}
\newcommand{\mjd}{{\small {MJD}}}
\newcommand{\adu}{{\small {ADU}}}
\newcommand{\hst}{{\it {\small {HST}}}}
\newcommand{\csp}{{\small {CSP}}}
\newcommand{\snf}{{\small {SNF}actory}}
\newcommand{\cfa}{{\small {CfA}}}
\newcommand{\loss}{{\small {LOSS}}}
\newcommand{\arc}{{\small {ARC}}}
\newcommand{\kpno}{{\small {KPNO}}}
\newcommand{\mdm}{{\small {MDM}}}
\newcommand{\wht}{{\small {WHT}}}
\newcommand{\het}{{\small {HET}}}
\newcommand{\ntt}{{\small {NTT}}}
\newcommand{\nnot}{{\small {NOT}}}
\newcommand{\salt}{{\small {SALT}}}
\newcommand{\vatt}{{\small {VATT}}}
\newcommand{\wiyn}{{\small {WIYN}}}
\newcommand{\signoi}{{\small {S/N}}}
\newcommand{\gri}{{\it {gri}}}
\newcommand{\ugriz}{{\it {ugriz}}}
\newcommand{\psf}{{\small {PSF}}}
\newcommand{\ra}{{\small {RA}}}

\newcommand{\raid}{{\small {RAID}}}
\newcommand{\pdf}{{\small {PDF}}}
\newcommand{\id}{{\small {ID}}}
\newcommand\fnu{$\rm{erg~cm}^{-2}~\rm{s}^{-1}~\rm{Hz}^{-1}$}
\newcommand\dme{\Delta m(B)}

\begin{document}

\title{The Sloan Digital Sky Survey-II Supernova Survey: Search Algorithm and
  Follow-up Observations}

\author{
Masao~Sako,\altaffilmark{1,2}
Bruce~Bassett,\altaffilmark{3,4}
Andrew~Becker,\altaffilmark{5}
David~Cinabro,\altaffilmark{6}
Fritz~DeJongh,\altaffilmark{7}
D.~L.~Depoy,\altaffilmark{8}
Ben~Dilday,\altaffilmark{9,10}
Mamoru~Doi,\altaffilmark{11}
Joshua~A.~Frieman,\altaffilmark{7,9,12}
Peter~M.~Garnavich,\altaffilmark{13}
Craig~J.~Hogan,\altaffilmark{14}
Jon~Holtzman,\altaffilmark{15}
Saurabh~Jha,\altaffilmark{2}
Richard~Kessler,\altaffilmark{9,16}
Kohki Konishi,\altaffilmark{17}
Hubert~Lampeitl,\altaffilmark{18}
John~Marriner,\altaffilmark{7}
Gajus~Miknaitis,\altaffilmark{7}
Robert~C.~Nichol,\altaffilmark{19}
Jose~Luis~Prieto,\altaffilmark{8}
Adam~G.~Riess,\altaffilmark{18,20}
Michael~W.~Richmond,\altaffilmark{21}
Roger~Romani,\altaffilmark{2,22}
Donald~P.~Schneider,\altaffilmark{23}
Mathew~Smith,\altaffilmark{19}
Mark~SubbaRao,\altaffilmark{12}
Naohiro~Takanashi,\altaffilmark{11}
Kouichi~Tokita,\altaffilmark{11}
Kurt~van~der~Heyden,\altaffilmark{4}
Naoki~Yasuda,\altaffilmark{17}
Chen~Zheng,\altaffilmark{2,22}
John~Barentine,\altaffilmark{24,25}
Howard~Brewington,\altaffilmark{25}
Changsu~Choi,\altaffilmark{26}
Jack~Dembicky,\altaffilmark{25}
Michael~Harnavek,\altaffilmark{25}
Yutaka~Ihara,\altaffilmark{27}
Myungshin~Im,\altaffilmark{26}
William~Ketzeback,\altaffilmark{25}
Scott~J.~Kleinman,\altaffilmark{25,28}
Jurek~Krzesi\'{n}ski,\altaffilmark{25,29}
Daniel~C.~Long,\altaffilmark{25}
Elena~Malanushenko,\altaffilmark{25}
Viktor~Malanushenko,\altaffilmark{25}
Russet~J.~McMillan,\altaffilmark{25}
Tomoki~Morokuma,\altaffilmark{27,30}
Atsuko~Nitta,\altaffilmark{25,31}
Kaike~Pan,\altaffilmark{25}
Gabrelle~Saurage,\altaffilmark{25}
Stephanie~A.~Snedden\altaffilmark{25}
}

\altaffiltext{1}{
  Department of Physics and Astronomy,
  University of Pennsylvania, 209 South 33rd Street,
  Philadelphia, PA 19104.
}
\altaffiltext{2}{
  Kavli Institute for Particle Astrophysics \& Cosmology, 
  Stanford University, P.O. Box 20450, MS29,
  Stanford, CA 94309.
}
\altaffiltext{3}{
  Department of Mathematics and Applied Mathematics,
  University of Cape Town, Rondebosch 7701, South Africa.
}
\altaffiltext{4}{
  South African Astronomical Observatory,
  P.O. Box 9, Observatory 7935, South Africa.
}
\altaffiltext{5}{
  Department of Astronomy,
  University of Washington, Box 351580, Seattle, WA 98195.
}
\altaffiltext{6}{
  Department of Physics, 
  Wayne State University, Detroit, MI 48202.
}
\altaffiltext{7}{
  Center for Particle Astrophysics, 
  Fermi National Accelerator Laboratory, P.O. Box 500, Batavia, IL 60510.
}
\altaffiltext{8}{
  Department of Astronomy,
  Ohio State University, 140 West 18th Avenue, Columbus, OH 43210-1173.
}
\altaffiltext{9}{
  Kavli Institute for Cosmological Physics, 
  The University of Chicago, 5640 South Ellis Avenue Chicago, IL 60637.
}
\altaffiltext{10}{
  Department of Physics, 
  University of Chicago, Chicago, IL 60637.
}
\altaffiltext{11}{
  Institute of Astronomy, Graduate School of Science,
  University of Tokyo 2-21-1, Osawa, Mitaka, Tokyo 181-0015, Japan.
}
\altaffiltext{12}{
  Department of Astronomy and Astrophysics,
  The University of Chicago, 5640 South Ellis Avenue, Chicago, IL 60637.
}
\altaffiltext{13}{
  University of Notre Dame, 225 Nieuwland Science, Notre Dame, IN 46556-5670.
}
\altaffiltext{14}{
  Department of Astronomy,
  University of Washington, Box 351580, Seattle, WA 98195.
}
\altaffiltext{15}{
  Department of Astronomy,
  MSC 4500,
  New Mexico State University, P.O. Box 30001, Las Cruces, NM 88003.
}
\altaffiltext{16}{
  Enrico Fermi Institute,
  University of Chicago, 5640 South Ellis Avenue, Chicago, IL 60637.
}
\altaffiltext{17}{
  Institute for Cosmic Ray Research,
  University of Tokyo, 5-1-5, Kashiwanoha, Kashiwa, Chiba, 277-8582, Japan.
}
\altaffiltext{18}{
  Space Telescope Science Institute,
  3700 San Martin Drive, Baltimore, MD 21218.
}
\altaffiltext{19}{
  Institute of Cosmology and Gravitation,
  Mercantile House,
  Hampshire Terrace, University of Portsmouth, Portsmouth PO1 2EG, UK.
}
\altaffiltext{20}{
  Department of Physics and Astronomy,
  Johns Hopkins University, 3400 North Charles Street, Baltimore, MD 21218.
}
\altaffiltext{21}{
  Physics Department,
  Rochester Institute of Technology,
  85 Lomb Memorial Drive, Rochester, NY 14623-5603.
}
\altaffiltext{22}{
  Department of Physics,
  Stanford University,
  382 Via Pueblo Mall, Stanford, CA 94305.
}
\altaffiltext{23}{
  Department of Astronomy and Astrophysics,
  The Pennsylvania State University,
  525 Davey Laboratory, University Park, PA 16802.
}
\altaffiltext{24}{
  Department of Astronomy,
  McDonald Observatory, University of Texas, Austin, TX 78712.
}
\altaffiltext{25}{
  Apache Point Observatory, P.O. Box 59, Sunspot, NM 88349.
}
\altaffiltext{26}{
  Department of Astronomy,
  Seoul National University, Seoul, South Korea.
}
\altaffiltext{27}{
  Institute of Astronomy, Graduate School of Science, University of Tokyo,
  2-21-1, Osawa, Mitaka, Tokyo 181-0015, Japan.
}
\altaffiltext{28}{
  Subaru Telescope,
  650 North A'ohoku Place, Hilo, HI 96720.
}
\altaffiltext{29}{
  Obserwatorium Astronomiczne na Suhorze, Akademia Pedagogicazna w Krakowie,
  ulica Podchor\c{a}\.zych 2, PL-30-084 Krak\'ow, Poland.
}
\altaffiltext{30}{
  National Observatory of Japan,
  2-21-1, Osawa, Mitaka, Tokyo 181-8588, Japan.
}
\altaffiltext{31}{
  Gemini Observatory,
  670 North A'ohoku Place, Hilo, HI 96720.
}

\received{}
\revised{}
\accepted{}

\slugcomment{Submitted to The Astronomical Journal}
\shorttitle{SDSS-II Supernova Survey: Search and Follow-up}
\shortauthors{Sako et al.}

\begin{abstract}

  The Sloan Digital Sky Survey-II Supernova Survey has identified a large
  number of new transient sources in a 300~deg$^2$ region along the celestial
  equator during its first two seasons of a three-season campaign.  Multi-band
  (\ugriz) light curves were measured for most of the sources, which include
  solar system objects, Galactic variable stars, active galactic nuclei,
  supernovae (\sne), and other astronomical transients.  The imaging survey is
  augmented by an extensive spectroscopic follow-up program to identify \sne,
  measure their redshifts, and study the physical conditions of the explosions
  and their environment through spectroscopic diagnostics.  During the survey,
  light curves are rapidly evaluated to provide an initial photometric type of
  the \sne, and a selected sample of sources are targeted for spectroscopic
  observations.  In the first two seasons, 476 sources were selected for
  spectroscopic observations, of which 403 were identified as \sne.  For the
  Type Ia \sne, the main driver for the Survey, our photometric typing and
  targeting efficiency is 90\%.  Only 6\% of the photometric \snia\ candidates
  were spectroscopically classified as non-\snia\ instead, and the remaining
  4\% resulted in low signal-to-noise, unclassified spectra.  This paper
  describes the search algorithm and the software, and the real-time
  processing of the \sdss\ imaging data.  We also present the details of the
  supernova candidate selection procedures and strategies for follow-up
  spectroscopic and imaging observations of the discovered sources.

\end{abstract}

\keywords{cosmology: observations --- methods: data analysis --- techniques:
  image processing --- supernovae: general --- surveys}

\section{Introduction}
\label{section_intro}

  Measurements of luminosity distances to Type Ia supernovae (\sne) have
  played a central role in cosmology, leading two independent groups to the
  remarkable discovery of an unknown, presently-dominant component of the
  universe, dark energy, and strong evidence for an accelerating universe
  \citep{riess98,perlmutter99}.  Current surveys that target high-redshift
  \sne\ from the ground -- the Canada-France-Hawaii Telescope Supernova Legacy
  Survey (\snls\footnote{http://www.cfht.hawaii.edu/SNLS}; \citealt{astier06})
  and the Equation of State: Supernovae Trace Cosmic Expansion
  (\essence\footnote{http://www.ctio.noao.edu/essence}; \citealt{wood-vasey07,
  miknaitis07}) -- and from space using the \hst\
  \citep{riess04a,riess07,barbary06} have substantially increased the sample of
  high-$z$ \sne, and have provided much-improved statistical constraints on
  the expansion history of the universe.

  The discovery of cosmic acceleration was made possible in part through
  extensive observations of nearby Type Ia \sne\ by the Cal\'an/Tololo
  Supernova Search \citep{hamuy93, hamuy96a, hamuy96b, hamuy96c} and by the
  \cfa\ follow-up program \citep{riess95, riess96a, riess99, jha06, jha07},
  and by studies pioneered by \citet{pskovskii77} and by \citet{phillips93} of
  the relationship between peak brightness and light curve decline rate
  \citep{hamuy96d, riess96b, phillips99}.  Current low-redshift \sn\ surveys
  and follow-up programs (Lick Observatory Supernova Search --
  \loss\footnote{http://astro.berkeley.edu/\~bait/kait.html};
  \citealt{filippenko01}, Carnegie Supernova Project --
  \csp\footnote{http://csp1.lco.cl/~cspuser1/PUB/CSP.html}; \citealt{hamuy06},
  Nearby Supernova Factory -- \snf\footnote{http://snfactory.lbl.gov/};
  \citealt{aldering02}, and the \cfa\ \sn\
  Group\footnote{http://cfa-www.harvard.edu/oir/Research/supernova/SNgroup.html}),
  are continuing to discover \sne\ and compile a large number of high-quality
  multicolor light curves as well as multi-epoch optical spectra of \sneia\ to
  expand the library of local training data used as ``templates''.  These
  high-quality data sets will be indispensable for calibrating the
  brightness-decline relation to high precision.  Obtaining and studying
  multi-epoch spectra are also important for computing improved
  $K$-corrections and minimizing systematic uncertainties \citep{kim96,
  nugent02, hsiao07}.  Recent spectroscopic modeling efforts have also led to
  a better understanding of the physical mechanism responsible for the
  observed brightness-decline relation (see, e.g., \citealt{kasen07} and
  references therein).

  As one of the three primary scientific components of the Sloan Digital Sky
  Survey-II (\sdssii), the Supernova Survey takes repeated imaging scans of
  the same 300~square degrees of the sky during the Fall seasons of 2005 --
  2007 to search for and measure light curves of \sne. The imaging survey is
  complemented by an extensive spectroscopic follow-up program to confirm the
  \sn\ type and measure redshifts, and to study the detailed spectral
  properties of a sample of selected events.

  This program exploits the unique capabilities of the \sdss\ 2.5m telescope
  \citep{gunn06} and its \ccd\ imaging camera \citep{gunn98} to survey a large
  volume of space at moderately high cadence.  The survey complements and
  improves upon other low-z and high-z surveys in several important ways.  The
  wide field-of-view camera operating in drift scan mode allows for efficient
  discoveries of type Ia \sne\ at $0.05 \la z \la 0.4$, a redshift interval
  that is not easily probed by other existing surveys that target either known
  nearby galaxies (low-$z$) or narrow pencil-beam volumes (high-$z$).  Its
  well-calibrated multi-band photometric system ($ugriz$;
  \citealt{fukugita96}) enables precise measurements of supernova light curves
  with controlled systematics.  The absolute magnitude scale is accurate to
  better than $\sim 2$\% in $r$ and $\sim 2 - 3$\% in the colors \citep{dr5},
  and a factor of $\sim 2$ improvement has been obtained from repeat imaging
  of the equatorial region \citep{ivezic07}.  Finally, the survey is sensitive
  to the redshift interval that, given a large enough sample, enables
  cosmological distance measurements with data from a {\it single telescope},
  eliminating the need for cross-calibration across two or more photometric
  systems.

  This paper is part of a series describing the \sdssii\ Supernova Survey.
  Here we present a technical description of the search algorithm, data
  processing, photometric typing of \sn\ candidates, and spectroscopic target
  selection.  \citet{frieman07} presents an overview of the program.
  Photometry and light curves of the full sample of spectroscopically
  confirmed \sne\ from the 2005 season are presented in \citet{holtzman07}.
  Spectroscopic data and their analysis results are described in
  \citet{zheng07}.  \citet{kessler07} presents the \snia\ Hubble diagram and
  cosmological analysis from the 2005 season.  The measurement of the
  low-redshift Type Ia \sn\ rate is presented in \citet{dilday07}.  Detailed
  studies of two peculiar \sne\ discovered by the \sdssii\ \sn\ Survey,
  \sn2005hk and \sn2005gj, are presented in \citet{phillips07} and
  \citet{prieto07}, respectively.

  The main body of the paper is separated into two broad sections.  The first
  part (\S\ref{sec:obs}) presents the details of the real-time on-mountain
  data processing and the mechanics of the search pipeline, which identifies
  new transient events.  The second part (\S\ref{sec:spectro}) describes the
  procedures for supernova candidate identification, photometric \sn\ typing,
  and the algorithm adopted for selecting targets for spectroscopy.  A brief
  discussion of follow-up imaging observations of a sample of
  spectroscopically confirmed \sne\ is presented in \S\ref{sec:imaging}.  The
  general results from the 2005 and 2006 seasons are presented in
  \S\ref{sec:results}.  We briefly summarize in \S\ref{sec:summary}.

\section{SDSS 2.5m Observations and Data Processing}
\label{sec:obs}

  The \sdssii\ Supernova Survey has been allocated the bulk of the time during
  the Fall seasons (September 1 - November 30) of 2005 -- 2007 on the
  \sdss\ telescope at Apache Point Observatory (\apo).  Imaging observations
  are scheduled on most nights excluding a 5-day period around full moon.
  Some nights are shared with the {\small SEGUE} program especially during
  stretches of consecutive nights with good observing conditions.  An overview
  of the \sdss\ is given by \citet{york00}; see \citet{frieman07} for an
  overview of the \sn\ Survey.

\subsection{Survey Area: Stripe 82}
\label{subsec:stripe82}

  The survey area is Stripe 82, the 300~deg$^2$ southern equatorial stripe of
  the \sdss\ footprint, which covers the approximate coordinate ranges
  $-60^\circ \la \alpha \la +60^\circ$ (20~hrs to 4~hrs in right ascension,
  \ra\ or $\alpha$) and $-1.25^\circ \la \delta \la +1.25^\circ$ in
  declination ($\delta$).  The detailed Stripe 82 footprint of each camera
  column is shown in Table~\ref{tbl:stripe82}.  This survey area was selected
  for three primary reasons: (1) extensive repeat observations were acquired
  as part of the \sdssi\ survey, before the start of the \sn\ Survey, (2) the
  area is easily accessible to most telescopes, and (3) it is accessible at
  low airmass during the Fall months, when most of the northern \sdss\ area is
  not.  The advantages of having repeat observations from \sdssi\ are
  three-fold.  First, the data provide a catalog of known variable sources,
  which is crucial for distinguishing a small number of new \sn\ candidates
  from a large population of foreground variable stars and background active
  galactic nuclei.  Second, the deep coadded images constructed from the
  individual scans serve as references for image subtraction, enabling a more
  sensitive search of new transient events (the template is essentially
  noiseless in the detection process).  Finally, the deep coadds also allow
  identification of faint host galaxies that are otherwise undetected in the
  single-scan images, which are frequently useful for prioritizing follow-up
  observations.  The stripe contains over 3 million cataloged galaxies that
  are brighter than $r \sim 22.5$~mag.

\begin{figure}[!tb]  
  \begin{center}
    \includegraphics[angle=0, width=.65\textwidth]{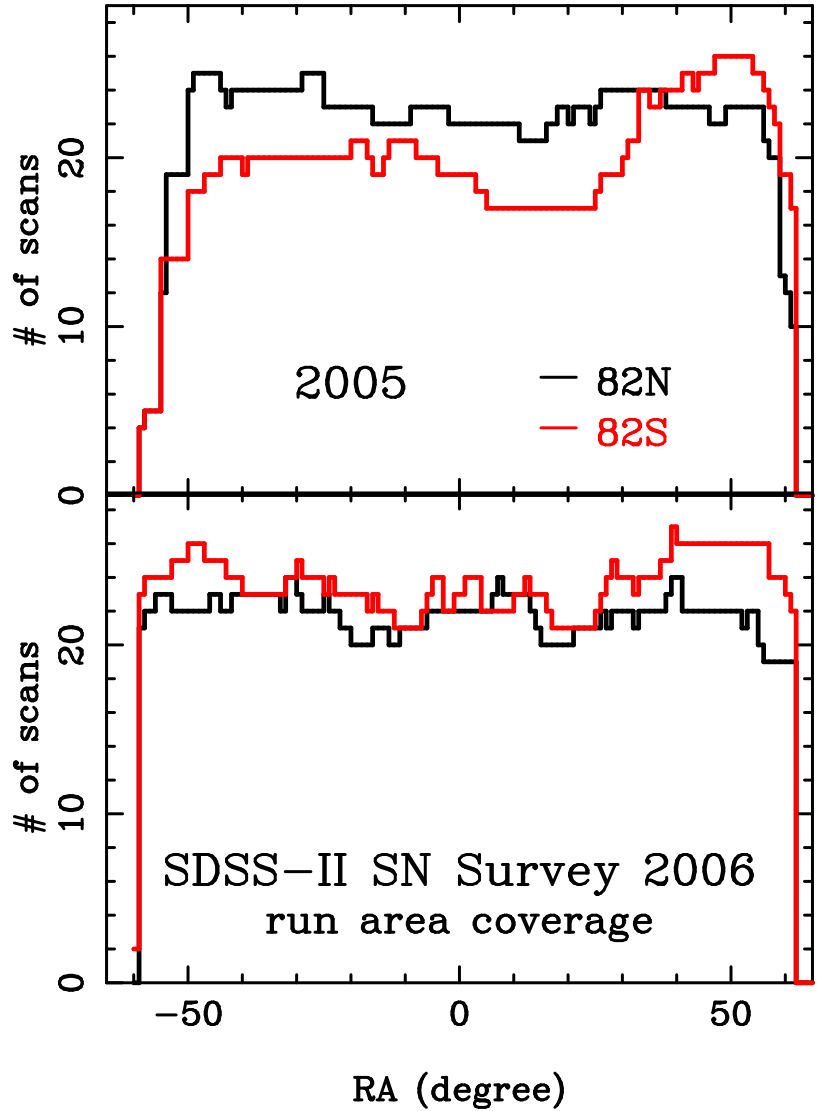}
  \end{center}
  \caption{Number of imaging scans of the northern (black) and southern (red)
    strips for the 2005 (top) and 2006 (bottom) seasons as a function of right
    ascension.  Note that, in contrast to the 2005 season, the 2006 scans are
    more evenly distributed in right ascension with very little difference
    between the northern and southern strips.} \label{fig:runarea}
\end{figure}

  Due to gaps between the six \ccd\ columns, the \sdss\ imaging camera is
  capable of scanning approximately half of the stripe (or one strip) in a
  single night.  During the 2005 season, the observations alternated between
  the northern strip (82N) and the southern strip (82S) from night to night.
  In 2006, however, a significant effort was made to avoid long temporal gaps
  in any given part of the strip, so pieces of \ra\ ranges from both strips
  were sometimes observed in a single night.  We also note that approximately
  10\% of Stripe 82 is covered by both the northern and southern strips due to
  overlapping \ccd\ columns, so $\sim 30$~deg$^2$ of the sky is observed on a
  cadence of 1 day (modulo weather losses).  The lists of all
  \sdssii\ \sn\ runs, i.e., continuous imaging scans, and their corresponding
  \ra\ ranges taken during the 2005 and 2006 seasons are given in
  Table~\ref{tbl:runs2005} and Table~\ref{tbl:runs2006}, respectively.  We
  also show in Figure~\ref{fig:runarea} the number of visits made to each of
  the strips in 2005 and 2006 as a function of the \ra.  The complete set of
  corrected frames and the uncalibrated object catalogs from 2005 and 2006 are
  available online as part of the first \sn\ data release ({\small
    DRSN}1\footnote{http://www.sdss.org/drsn1/DRSN1\_data\_release.html}).
  These data are also accessible from the Data Archive Server of the Sixth
  Data Release\footnote{http://www.sdss.org/dr6/access/index.html}
  \citep{dr6}.

\subsection{On-Mountain Processing}
\label{subsec:processing}
 
  The \sdss\ 2.5m telescope nominally performs observations every night during
  the search seasons, so the data must be processed within 24~hours to avoid a
  backlog of unprocessed data and to rapidly identify \sn\ candidates for
  spectroscopic follow-up observations.  A single night of observing produces
  up to $\sim 200$~Gb of imaging data, so it is highly desirable to process
  the data on the mountain and significantly reduce the number of images that
  are transferred over the internet for visual inspection.

  The data are processed on a dedicated 10 dual-processor (20 processors
  total) computer cluster that runs at \apo.  There are nine computers that
  process data; each of the 18 processors is responsible for processing all
  images from a single combination of filter (\gri) and camera column (1 --
  6).  Each computer has 400~Gb of \raid 5 disk and 2~Gb memory.  The 10th
  monitor/backup computer has 3~Tb of disk and 4~Gb memory and is used for
  process-management, data transfer, and software development.  This setup was
  specifically chosen to meet our requirement of processing a full night's
  worth of data within 24~hours.  An 11th computer is available as a spare but
  so far has not been needed -- during the first two years of operation, there
  were no serious failures.  Several disks failed, but the \raid\ system
  worked properly to prevent any loss of processing time while waiting for a
  disk to be replaced.  There was one curious glitch that may be related to
  operation at high altitude (2800~m).  On occasion one of the computers would
  hang up, and cycling the power was the only way to revive the system.  There
  were about 20 such incidents per season, although the probability seemed
  higher in the month of November, possibly due to better weather and longer
  nights resulting in more computing time.  Since people are present at
  \apo\ day and night, minimal processing time was lost from these hang-ups.

  There are three main tasks for the process manager: (i) start jobs, (ii)
  monitor processing progress, and (iii) monitor disk space.  Due to the
  structure of the data acquisition system, the image processing can start
  only after an imaging run has finished, and the master script is usually
  executed manually by a person in the morning.  While this script could have
  been automatically started, it is better that a human reviews the observer
  logs and checks that processing starts smoothly.  The master script
  schedules and allocates resources for copying data from the data acquisition
  computer to a local disk, photometric reduction in $ugriz$
  (\S~\ref{subsub:photo}), and frame subtraction (\S~\ref{subsub:framesub})
  and object detection (\S~\ref{subsub:doobjects}) in $gri$.  Several monitor
  scripts are used to check the status throughout the $\sim 20$~hours needed
  to process the data.  Since data processing continues virtually
  round-the-clock, a few people often shared the monitoring burden.
  Continuous monitoring was necessary because there were two common problems
  that could interupt the processing.  First, poor observing conditions caused
  the photometric reduction software to abort.  In this case, the photometric
  reduction must be reprocessed with a more restrictive \ra\ range.  The
  second source of interruption was the computer hang-ups discussed above.
  The last process-manager issue concerns disk space.  Rather than clearing
  disk space after each night's data have been processed, we kept all of the
  subtracted images at \apo\ for 1--2 weeks.  The reason for keeping
  subtracted images is that for interesting candidates we could go back to
  earlier epochs and process $u$ and $z$: $u$ band was particularly useful to
  distinguish Type Ia and Type II supernovae.  In addition, since the
  candidate position is known, we re-ran $gri$ photometry at the known
  location in previous epochs to get a better estimate of the early-epoch
  fluxes and upper limits.  The disk space was managed with a priority system
  so that clearing disk was mostly based on following a pre-defined algorithm
  and rarely involved a hasty decision.  To avoid risks with automated
  disk-cleaning, a human-entered command was required.

  Below we describe the search pipeline and the reduction procedures adopted
  to take the raw images to the point where they are transferred to the central
  \sn\ database server at Fermilab.

\subsubsection{PHOTO}
\label{subsub:photo}

\begin{figure}[tb]  
  \begin{center}
    \includegraphics[angle=270, width=.65\textwidth]{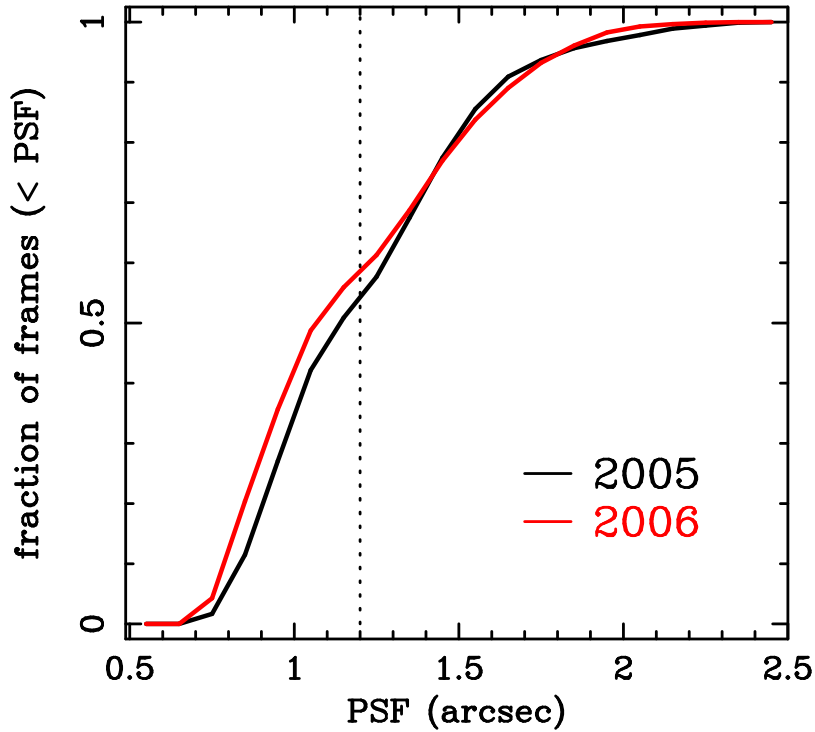}
  \end{center}
   \caption{Cumulative distributions of the $r$-band \psf\ during the 2005
     (black) and 2006 (red) search seasons.  More than half of the frames were
     acquired at $< 1.2$\arcsec.  Only frames successfully processed through
     \photo\ are included.} \label{fig:seeing}
\end{figure}

  The raw data are first processed through the \photo\ pipeline
  \citep{lupton01,lupton02,stoughton02}.  The software produces the corrected
  frames and generates bad-pixel maps, position-dependent \psf, and
  astrometric solution \citep{pier03} that are used in subsequent processing
  stages.  Our version of \photo\ does not identify sources in the images or
  perform photometric measurements \citep{hogg01,smith02,ivezic04,tucker06},
  as they are not appropriate for efficiently identifying \sn\ candidates,
  which are usually blended with their host galaxies.  As with all standard
  \sdss\ data, images from each camera column are written in frames of $2048
  \times 1361$ unique pixels or 13.51\arcmin\ $\times~8.98$\arcmin\ on the
  sky.  The pipeline software was slightly tweaked to run on data taken under
  poor observing conditions (bright moon, low atmospheric transparency, and
  poor seeing).  In 2005 and 2006, approximately 34\% and 30\%, respectively,
  of the frames were acquired when the moon was above the horizon.
  Figure~\ref{fig:seeing} shows the distributions of the $r$-band \psf\ of
  frames successfully processed through \photo\ during the first two seasons.

\subsubsection{Image Subtraction}
\label{subsub:framesub}

  To identify new transients in the search data, the images are run through a
  difference imaging pipeline, which matches the \psf\ of the search and
  template frames, adjusts for the difference in photometric zeropoint,
  accurately registers their pixels, and performs the subtraction.  The image
  templates were divided into four separate \ra\ ranges so that human scanning
  and target selection could begin well before the entire night of data was
  finished processing.  The subtracted frames in each filter are searched for
  positive fluctuations that consist of at least 2 contiguous pixels each
  above $3.0~\sigma$ of the noise.  The software is derived from version 7.1
  of the {\tt Photpipe} software used by the Super{\small MACHO} and \essence\
  collaborations \citep{smith02}.  Extensive changes have been implemented to
  operate successfully on processed \sdss\ data.  The modified application is
  called {\tt Framesub}.

  The pipeline is operated in three modes: 1) {\tt sdssred} mode, where the
  input of images is prepared for difference imaging.  2) {\tt sdssdiff},
  which performs a difference imaging analysis of the search data compared to
  the template data.  To conserve computing time, only the \gri\ frames are
  differenced and used for object detection.  This is the default mode that
  produces the objects for manual inspection, and subsequently the \sn\
  candidates described below.  3) {\tt sdssforce} mode, in which, after
  reliable detection of a supernova candidate, we difference {\it all} data on
  the object, including the $u$ and $z$ bands, and perform forced-positional
  photometry at the location of the candidate.  The measurements of magnitudes
  by {\tt sdssdiff} and {\tt sdssforce} are referred to as {\it search} and
  {\it forced} photometry.  The overall structure of the software and detailed
  descriptions of each mode are provided in Appendix~\ref{app:software}.

\subsubsection{Object Detection and Filtering Algorithm}
\label{subsub:doobjects}

  The individual peaks found by {\tt Framesub} are filtered through a software
  called {\tt doObjects} to remove statistical fluctuations and identify true
  astronomical sources.  First, the single-filter peaks are matched by
  position to identify sources that are detected in at least two of the three
  \gri\ filters within $0.8$\arcsec.  These sources are flagged as {\it
  objects}.  Significant negative fluctuations are not flagged.  The list of
  objects is compared against the list of known variable sources (the veto
  catalog) constructed from previous scans of Stripe 82.  Any object that
  matches the position of a known variable is filtered out.  Most of the
  sources in this catalog are variable stars, active galactic nuclei (\agn),
  and other persistently varying sources.  Approximately 10\% (20\%) of the
  objects identified by {\tt doObjects} in 2005 (2006) were associated with
  sources in the veto catalog.

  Given the large area covered by the survey and the overlap of the central
  \ra\ range of Stripe~82 with the ecliptic plane, our detections are
  overwhelmingly dominated by solar system objects.  The challenge is to
  remove as many of these objects as possible prior to handscanning, without
  filtering out real \sn\ candidates.  Objects with proper motions in excess
  of $\sim 1$\arcsec\ per minute are easily rejected by requiring that the
  object is detected in the $r$ and $i$ bands within $0.8$\arcsec\ (the
  filters are imaged in the order $riuzg$ and the effective time difference
  between adjacent filters is 71.7 seconds).  Objects that move as slowly as
  $\sim 0.2$\arcsec\ per minute can be identified through motion between the
  $g$ and $r$ bands.  Objects that move at a slower rate are slightly more
  difficult to identify; when they pass in front of a background galaxy, they
  can look like perfectly good \sn\ candidates.  Approximately 35\% (40\%) of
  the objects found in 2005 (2006) were tagged as moving objects and removed
  prior to human evaluation.

  New transient sources are entered into a dedicated {\tt MySQL} database and
  the cutout images are transferred to Fermilab for visual inspection.

\subsection{Handscanning, Autoscanning, and SN Candidate Selection}
\label{subsec:handscan}

  In addition to epochs of \sn\ light-curves, the difference imaging and
  object detection algorithms described above result in the detection of many
  background sources of variation, both physical and non-physical.  In order
  to robustly reject background and allow us to focus further analysis on
  promising \sn\ candidates, we require that a human visually inspect images
  of the objects, a process which we refer to as {\it handscanning}.

\begin{figure}[!tb]
  \begin{center}
  \includegraphics[angle=0, width=1.00\textwidth]{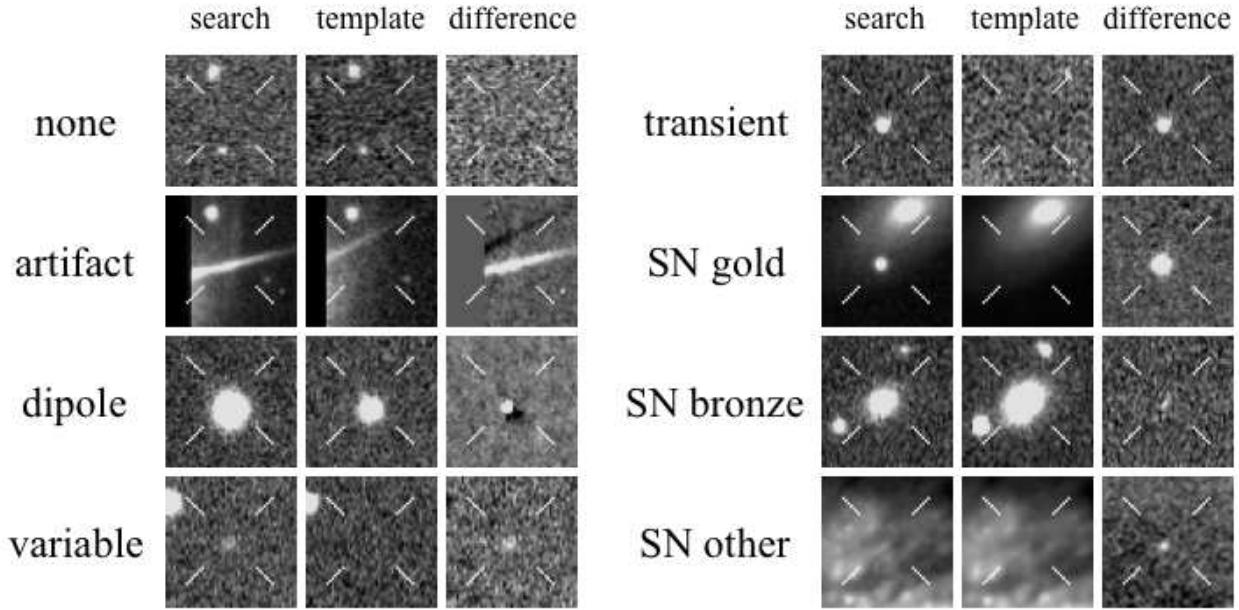}
  \end{center}
  \caption{A gallery of various types of objects that a scanner evaluates.
    The panels show $r$-band search (left columns in each panel), template
    (central columns), and differenced (right columns) images of objects
    classified by scanners as ``none'', ``artifact'', ``dipole'',
    ``variable'', ``transient'', ``\sn\ gold'', ``\sn\ bronze'', and
    ``\sn\ other''.  Objects classified as ``none'' are typically detections
    near the threshold, and are often not visible to the human eye.  Most
    ``artifacts'' are caused by different orientations of diffraction spikes
    in the search and template images.  A ``dipole'' results from imperfect
    image registration as well as high proper-motion stars.  A ``variable'' is
    an object associated with a persistent source and usually exhibits random
    temporal variability.  These sources are newly variable sources that are
    not part of the veto catalog.  A ``transient'' object is usually seen for
    the first time and {\it not} associated with a galaxy.  The source can be
    a true \sn\ with an undetected host galaxy or a slow-moving asteroid.  An
    ``\sn\ gold'' is a good \sn\ candidate that is well-separated from the
    host.  An ``\sn\ bronze'' is a possible \sn\ that lies close to the core
    of the galaxy and can, therefore, also be a variable active galactic
    nucleus.  An ``\sn\ other'' is also a possible \sn, but is most likely a
    different type of variable source -- in the case shown above, the source
    is a variable star in NGC1068 (M77).  There is another category called
    ``\sn\ silver'' (not shown), which is used for objects that are similar to
    a ``transient'', but detected in more than a single epoch (e.g., an
    \sn\ with an undetected host galaxy).  The distinction between
    gold/silver/bronze/other \sn\ candidates is not important; all objects
    tagged as an ``\sn'' become a \sn\ candidate.  Each cutout image is
    20\arcsec\ on the side.}
\label{fig:object_mosaic}
\end{figure}

\begin{figure}[tb]
  \begin{center}
  \includegraphics[angle=0, width=0.4\textwidth]{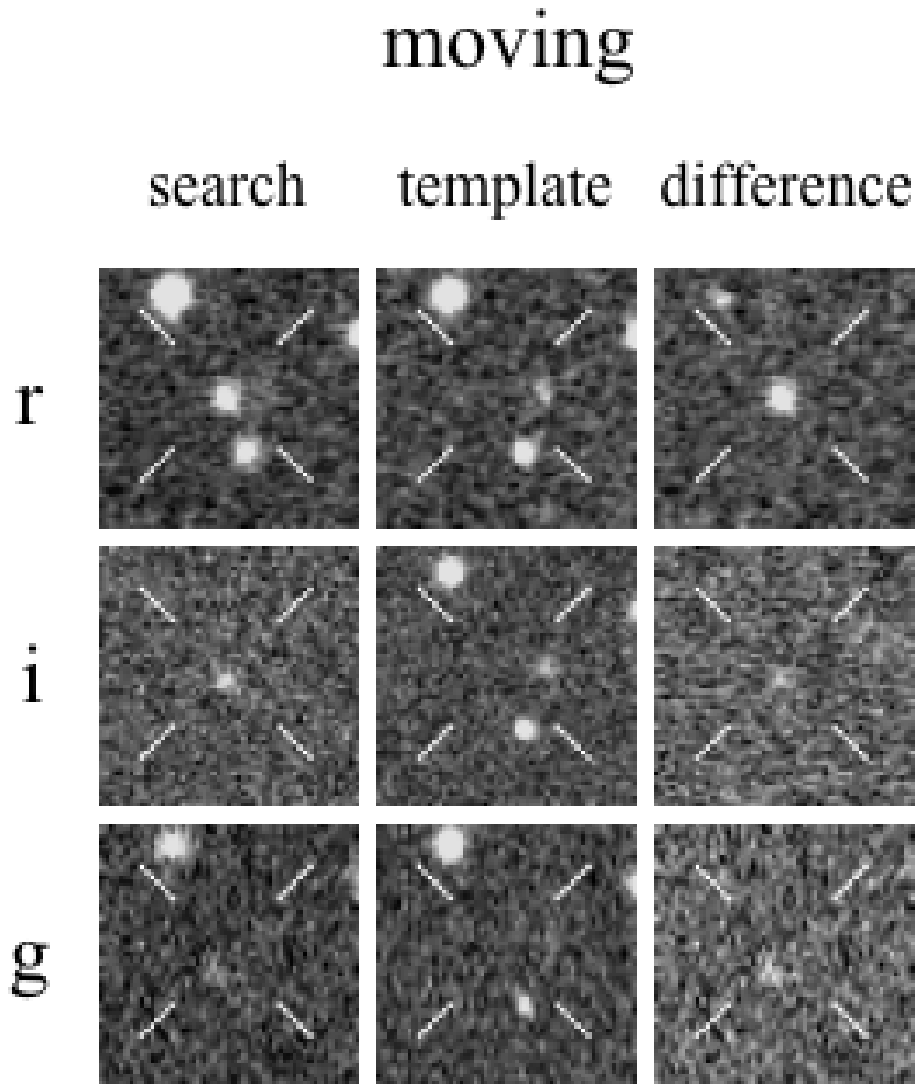}
  \end{center}
  \caption{An example of a moving object that was correctly flagged by the
    autoscanner.  As in Figure~\ref{fig:object_mosaic}, the search, template,
    and differenced images are shown in the left, middle, and right columns,
    respectively.  In addition to the $r$-band images (top row), we also
    display the $g$ (middle row) and $i$-band (bottom row) images to show the
    relative position of the source in each filter.  Note that the faint
    source in the $g$-band differenced image appears slightly below and to the
    left of the center of the image.  The images are acquired in the order
    $ri(uz)g$ with a 71.7-second delay between each filter.}
\label{fig:object_moving}
\end{figure}

  In addition to moving solar system objects, other major sources of
  contamination include artifacts caused by (1) subtraction of slightly
  misaligned images, which creates shapes with clusters of positive and
  negative counts (objects that we call dipoles) in the differenced images,
  (2) diffraction spikes, and (3) bright saturated stars.  Satellite trails
  that are not properly masked by the software also contribute to the
  background.  A small fraction of the dipoles are high proper-motion stars
  that have drifted between the epochs of the template and search images.
  Many of these background objects, however, can be quickly rejected by visual
  inspection.  In Figure~\ref{fig:object_mosaic}, we show a gallery of cutout
  search, template, and differenced images of various types of objects
  evaluated by scanners -- none, artifact, dipole, variable, transient,
  \sn\ gold, \sn\ bronze, and \sn\ other.  All images shown are from the $r$
  band.  In Figure~\ref{fig:object_moving}, we show \gri\ images of a moving
  object correctly flagged by the software.  These images show the source
  moved between the $g$ and $r$ exposures, which are separated by about 5
  minutes.

  As a convenient way for humans to handscan the detected objects, a web
  interface was constructed that queries the database and displays all of the
  relevant information about the object, including \gri\ cutout images of the
  search, template, and differenced images, measured magnitudes, the mean sky
  coordinates and the relative positions measured in each filter, and a list
  of all previous detections (if any) within 0.8$\arcsec$ of the object under
  inspection.  The scanner evaluates the information and decides to either tag
  the object as a possible \sn\ or to reject it as background.  If an object
  is tagged as a \sn\ and there are no previous detections of objects within
  0.8$\arcsec$, it becomes a new \sn\ {\it candidate} with a unique
  \sn\ {\small ID} number.  A candidate will always remain a candidate unless
  it is manually vetoed by one of the scanners (which rarely occurs).  All of
  the candidates can be accessed through a public web
  site\footnote{http://sdssdp47.fnal.gov/sdsssn/candidates/candTable.php}.

  During the 2005 observing season, we required that every object that is not
  rejected by {\tt doObjects} be handscanned by a human.  An average of about
  3000 -- 5000 objects were inspected per full night of imaging during the
  Fall 2005 season.  Six scanners were on duty on a given night, each scanner
  responsible for inspecting all objects from one of the six camera columns,
  or 500 -- 800 objects.  Based on our experience in 2005, we made a number of
  changes to the handscanning procedure for the 2006 season described below.

\subsubsection{Autoscanner}
\label{subsec:autoscan}

  To reduce the number of objects to be scanned by humans, prior to the 2006
  season we implemented a new software filter, called the {\it autoscanner}.
  The software performs two primary tasks: 1) identifies all objects detected
  in more than one epoch as well as bright ($g$ or $r<21$~mag) objects
  detected for the first time, and 2) uses statistical classification
  techniques to identify and filter out first-epoch background non-\sn\
  objects.

  The reasons for performing 1) are as follows.  First, the selection of
  objects detected in more than one epoch provides a very robust way of
  eliminating moving objects, one of the major contaminants of the 2005 \sn\
  candidates.  Second, the selection of bright first-epoch objects enables us
  to discover nearby \sne; we can thereby obtain spectroscopic observations
  well before maximum light and provide rapid alerts to the \sn\ community.
  In addition, bright, single-epoch candidates provide a back-up list of
  spectroscopic targets for nights when the spectroscopy queue is not filled
  by promising multi-epoch candidates.  Although most of our targets were
  spectroscopically observed after two or more epochs of photometry, in a few
  instances, we did obtain spectroscopic observations of \sn\ candidates based
  on a single epoch of detection.  In the selection of these objects, the
  autoscanner also identifies objects associated with either a known variable
  in the veto catalog or a known \sn\ candidate.  In the former case, the
  object is filtered out.  In the latter case, since there is no need for a
  human to scan the object, it is instead flagged by the autoscanner as a
  known \sn\ candidate, stored in the database, and used in the light curve
  analysis (see Section~\ref{subsec:snphoto}).

  For 2), the autoscanner attempts to identify and filter out bright
  first-epoch objects that belong to one of three classes of background;
  unmasked diffraction spikes, artifacts of imperfect image registration
  (dipoles), and moving objects.  The software treats the set of objects,
  along with their evaluations by a human, from the 2005 search as a training
  set, and compares the observed quantities for newly detected objects against
  this set.  The method used to classify an object is the {\it histogram
  method of probability density function (\pdf) estimation}.  In this method
  the observable quantities, or attributes, of an object form a
  multi-dimensional space, and a classification decision is made by
  considering the number count of objects, in the training set, from each
  class in a bin centered at the point describing the object we wish to
  classify.  This method has the advantages of being non-parametric, allowing
  the decision boundaries in the observable space to be arbitrarily
  complicated, and of explicitly retaining the correlations between
  observables.  A caveat of using the histogram method of \pdf\ estimation is
  that one must have a sufficient number of objects in the training set to
  sample the \pdf\ well in all regions of interest.  With $\approx 90,000$
  objects in the training set, this is not a significant limitation for the
  \sdssii\ \sn\ Survey.  A technical discussion of the algorithm is presented
  in Appendix~\ref{app:autoscanner}.

  During the 2006 observing season, the autoscanner was used to reject moving
  objects from among the set of bright first-epoch objects that were to be
  handscanned.  For such bright objects, only $\approx 0.7$\% of them are
  incorrectly tagged as moving objects by the autoscanner.  In contrast to the
  2005 observing season, we chose not to handscan epochs of known candidates
  or objects that showed variation in more than one observing season, and the
  autoscanner was used to reject such objects.  Although some objects were
  classified as artifacts or dipoles, we chose not to reject these from the
  handscanning as they are a small fraction of the background in comparison to
  moving objects, and the relatively small reduction in handscanning that
  would be accomplished was determined to be outweighed by the risk of missing
  an early epoch of a nearby \sn.

  This new handscanning strategy and the autoscanner enabled a reduction in
  the number of objects scanned by more than an order of magnitude between
  2005 and 2006, with no reduction in the quality or quantity of confirmed
  \sne.  While scanning a single camera column in 2005 typically took $2 -
  3$~hours per person for a full night of data, in 2006 a scanner could cover
  two columns in only $10 - 20$~minutes.  The reduction of the 2006
  handscanning by the autoscanner is summarized in Table~\ref{tbl:scanning}.

  A comparison of the objects classified by the autoscanner as artifacts to
  their human classification shows that the autoscanner is extremely efficient
  at recognizing artifacts, and that our concern about potentially rejecting
  nearby \sne\ was overly cautious.  During the 2006 observing campaign, 3753
  objects had been classified by the autoscanner as artifacts.  Humans
  classified 2668 of these objects as artifacts and 3710 of them as some type
  of non-\sn\ background.  Of the remaining 43 objects classified by a humans
  as \sn\ candidates, 41 were background (non-\sn) erroneously classified by
  the scanner, one was a bright nova outburst, and one was the first epoch of
  a \sn\ that was later (after the 2nd epoch detection) classified as a \sn.
  Therefore, in cases where they disagree on classification of artifacts, the
  autoscanner appears to be more reliable than the human scanners.

\section{Spectroscopic Target Selection and Follow-up Observations}
\label{sec:spectro}

  Spectroscopic follow-up is an essential component of any current supernova
  survey, for both confirming the type of the \sn\ candidate and measuring the
  cosmological redshift (preferably from host galaxy emission/absorption
  lines).  For the \sdssii\ \sn\ survey, the amount of scheduled spectroscopic
  telescope time is larger than the amount of imaging time on the \sdss\ 2.5m
  telescope by a large factor (see also \citealt{frieman07}).  The telescopes
  used for spectroscopy in 2005 and 2006 included the 9.2m \het, 3.6m \ntt,
  3.5m \arc, 8.2m Subaru, 2.4m \mdm\ Hiltner, 4.2m \wht, 4m \kpno\ Mayall, 10m
  Keck, 2.5m \nnot, and the 11m \salt.

  On any given night during the fall season, there could be several telescopes
  that are simultaneously scheduled and observing a set of our \sn\
  candidates.  This requires a significant amount of coordination between the
  various observers at each of the telescopes to avoid observing the same
  candidate.  A single person (the first author of this paper) was responsible
  for coordinating all follow-up imaging and spectroscopic observations in
  2005 and 2006.  Telescopes separated by several time zones are relatively
  easy to coordinate, but those that have similar longitudes (e.g., \mdm,
  \arc, \kpno) require real-time communication.  In general, the brighter
  targets are parsed to telescopes with smaller aperture, but the magnitude
  limits must be adjusted depending on the observing conditions at each site.
  We have made every effort to avoid taking duplicate spectra, but on occasion
  the same object was observed nearly simultaneously on two telescopes.

  A good night of imaging of half of Stripe 82 typically yields $100 - 300$
  objects that are tagged as {\it new} \sn\ candidates, which are publicly
  accessible through the
  web\footnote{http://sdssdp47.fnal.gov/sdsssn/candidates/candTable.php}
  immediately after they are entered into the database.  This number
  significantly exceeds the number of targets we can observe
  spectroscopically, and choices must be made.  The following subsections
  describe the further prioritization of \sn\ targets for spectroscopic
  observations.  The algorithm and strategy depends on the amount of resources
  available, which is something that evolved quite substantially between the
  2005 and 2006 search seasons.

\subsection{SN Photometric Typing}
\label{subsec:snphoto}

  After each night of imaging on the \sdss\ 2.5m telescope, the \gri\ light
  curves of all active candidates are compared against a library of light
  curve templates of different \sn\ types.  The purpose of this procedure is
  twofold -- 1) to quickly identify the best-matching template and provide
  estimates of the redshift, extinction, the approximate date of maximum
  light, and current apparent magnitudes, and 2) to compute other quantities
  that are useful for prioritizing follow-up spectroscopy, such as the amount
  of host galaxy contamination.  The process is essential since the number of
  active \sn\ candidates (a good fraction of which are not \sne) at any given
  time exceeds the number that can be observed spectroscopically by a large
  factor.  A reliable system is necessary to make efficient use of the
  valuable spectroscopic resources.  Several techniques and algorithms for
  \sn\ photometric identification have been introduced previously by
  \citet{poznanski02, poznanski06}, \citet{johnson06}, \citet{riess04b},
  \citet{sullivan06}, \citet{kunz07}, and \citet{kuznetsova07}.  The method
  adopted here is most similar to that used by the \snls\ spectroscopic follow
  up described in \citet{sullivan06}.  Below, we describe the details of the
  algorithm.

  The measured light curves are compared against a library of light curve
  templates, which are grouped into three types -- \snia, \snibc, and \snii --
  and are generated from multi-epoch spectra constructed and compiled by
  P. Nugent\footnote{http://supernova.lbl.gov/~nugent/nugent\_templates.html}
  and also from the \suspect\
  database\footnote{http://bruford.nhn.ou.edu/~suspect/index1.html}.
  Specifically, for the Type Ia, we use Nugent's Branch-normal, 1991T-like,
  and 1991bg-like templates.  The Branch-normal spectra are used for computing
  synthetic light curves of \sneia\ with a range of luminosities parameterized
  by the decline rate.  The model is described in Appendix~\ref{app:snia_lc}.
  The set of templates used for the Type Ib/c are Nugent's normal and
  hypernova Ib/c spectra, as well as spectra and light curves of \sn 1999ex
  and \sn 2002ap from the \suspect\ database.  Similarly, we use Nugent's
  II-P, II-L, and IIn spectra, and \suspect 's 1993J (IIb), 1998S (IIn), and
  \sn 1999em (II-P) to generate a set of Type II light curves.

  The light curves in the observed \ugriz\ filters are calculated on a grid of
  four parameters ($z$, $A_V$, $T_{\rm{max}}$, [\dmB, template \sn]), where
  $z$ is the redshift, $A_V$ is the host galaxy extinction in the $V$ band and
  assumes $R_V = 3.1$, and $T_{\rm{max}}$ is the time of rest-frame $B$-band
  maximum light.  The last parameter refers to either the decline rate
  parameter for the Branch-normal Ia models (\dmB), or the particular
  \sn\ template for the peculiar \sneia\ (1991T-like and 1991bg-like) and the
  core-collapse models.  We do not attempt to fit or correct for the Milky Way
  extinction.  In this procedure, we {\it assume} a cosmology to convert the
  redshift to a luminosity distance, similar to the method adopted by the
  \snls\ described in \citet{sullivan06}.  The adopted cosmological parameters
  are $\Omega_m = 0.3$ and $\Omega_\Lambda = 0.7$.  For the \snia\ templates,
  we also assume a fiducial peak $B$-band absolute magnitude of $M_B = -19.0 +
  5 \log (H_0/70)$~mag, where $H_0$ is the Hubble constant in units of
  km/s/Mpc, for a standard \dmB = 1.1 \snia.  As shown by \citet{sullivan06},
  a particular set of assumed cosmological parameters in the computation of
  the model light curves does not significantly bias the population of targets
  for spectroscopic observations.  At low redshift ($z \la 0.2$), the
  luminosity distances are not sensitive to our choice of $\Omega_m$ and
  $\Omega_\Lambda$.  Above $z \sim 0.2$, the statistical uncertainties in the
  fluxes and the combination of varying $z$ and $A_V$ can compensate for
  differences in luminosity distance that might result from a different set of
  cosmological parameters.  The assumption, however, could systematically bias
  the estimated photometric redshifts, but that is not a concern for the
  purposes of target selection.

  The templates are grouped into three \sn\ types -- Ia, Ibc, and II -- and
  the fitter records the best-fit parameters and the minimum value of the
  $\chi^2$ statistic on the four-dimensional parameter grid within each of the
  three types.  In addition to identifying the \sn\ type with the lowest
  $\chi^2$ value, which we refer to as ``type-best'', we examine the relative
  values of the $\chi^2$ and determine whether the \sn\ candidate should be
  considered as ``typed'' according to one or both of the following two
  criteria.  If we denote the value of the $\chi^2$ of the best-matching type
  by $\chi^2_{\rm{min}}$, the next best type as $\chi^2_1$, and the worst one
  as $\chi^2_2$ (i.e., $\chi^2_{\rm{min}} < \chi^2_1 < \chi^2_2$), the ``A''
  criterion is satisfied if $\chi^2_{\rm{min}} < \chi^2_1 -
  \chi^2_{\rm{min}}$.  If ``type-best'' is a Ia {\it and} satisfies the above
  criterion, the candidate is said to have a ``type-A of Ia''.  The ``B''
  criterion is satisfied if $\chi^2_{\rm{min}} < N (\chi^2_{\rm{min}} +
  \chi^2_1 + \chi^2_2)/3$ with $N = 0.5$.  Similarly, if the best-fit type is
  a \snia\ and satisfies the above $\chi^2$ criterion, the candidate is said
  to have a ``type-B of Ia''.  The value of 0.5 adopted for $N$ works well for
  \sn\ candidates with low \signoi, and was empirically determined from a
  sample of spectroscopically confirmed \sne\ from our 2004 engineering run
  \citep{sako04}.  Also computed are the estimated current $g$ and $r$
  magnitudes from the model light curves.  We also search for the nearest
  galaxy within $10\arcsec$ from the \sn\ position in the \sdss\ galaxy
  catalog and refit and retype the light curves using the best estimate of its
  redshift as a prior.  We adopt galaxy photometric redshifts from
  \citet{oyaizu07} and spectroscopic redshifts from the \sdss\ {\small DR}5
  \citep{dr5}.

\begin{figure}[!tb]  
  \begin{center}
    \includegraphics[angle=270, width=.7\textwidth]{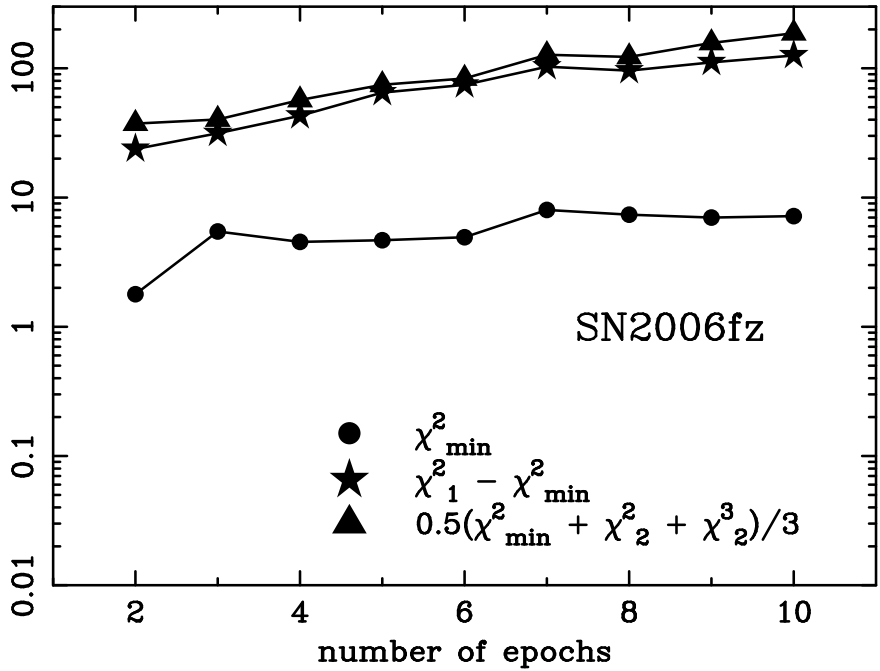}
  \end{center}
   \caption{Fit results for \sn2006fz (\snia\ at z=0.105; internal candidate
     \id\ 13135) as functions of the number of epochs in the light curve.  The
     bottom curve (labeled with filled circles) shows the minimum $\chi^2$
     values, which at every epoch correspond to the Ia model.  The top two
     curves labeled with filled stars and triangles show values relevant for
     the ``A'' and ``B'' typing criteria, respectively.  For this particular
     \sn, $\chi^2_{\rm{min}}$ is always below these two curves, which implies
     that this \sn\ is a Type Ia at high-confidence.}
   \label{fig:sn13135_chisq}
\end{figure}

\begin{figure}[!tb]  
  \begin{center}
    \includegraphics[angle=0, width=1\textwidth]{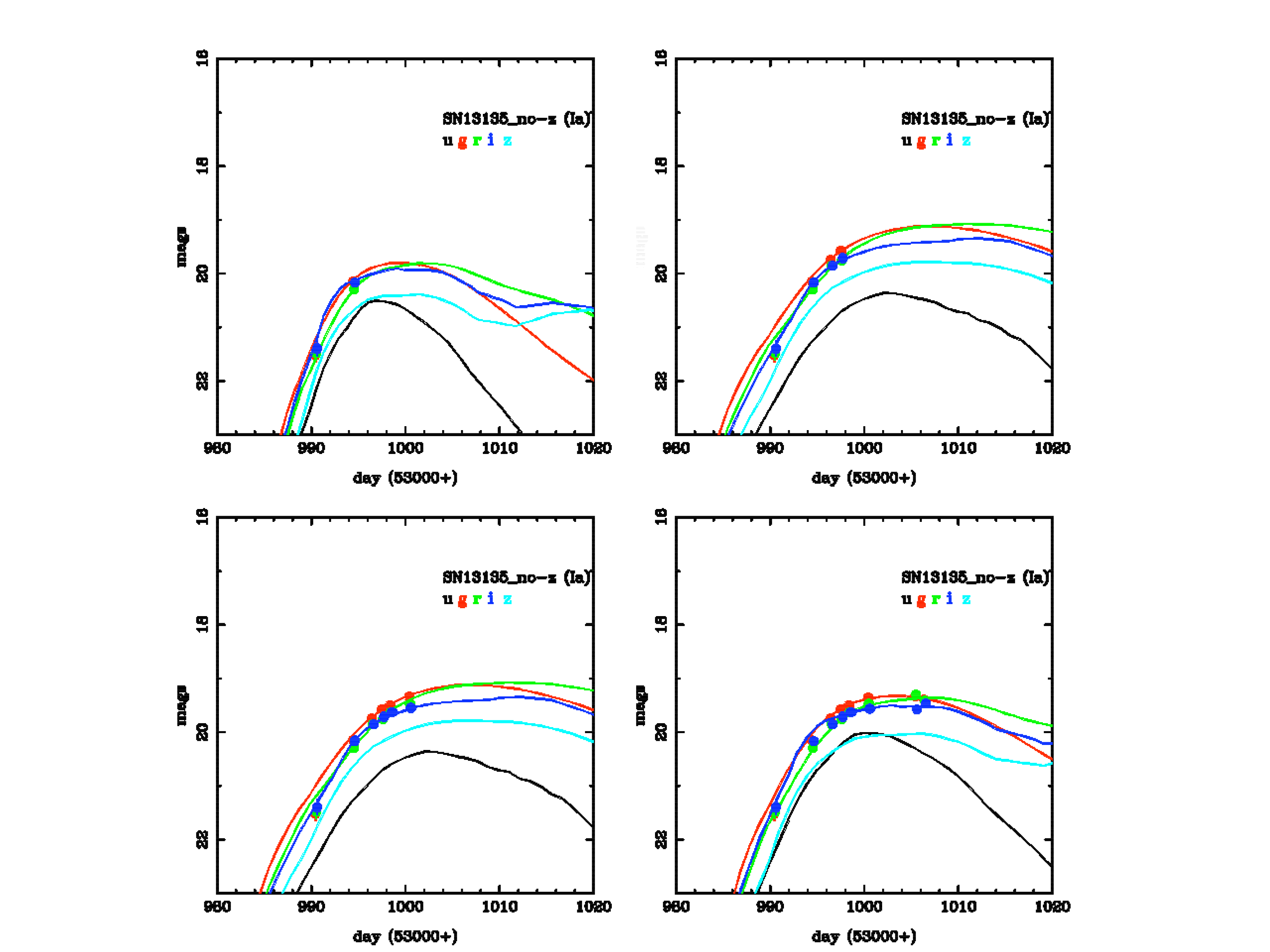}
  \end{center}
   \caption{Light curves of \sn2006fz (same \sn\ as shown in
     Figure~\ref{fig:sn13135_chisq}) and fits to the best-fit \snia\ models
     after 2, 4, 6, and 8 epochs of detections without a galaxy photometric
     redshift prior.  Search photometry from \gri\ are shown.  Model light
     curves in $u$ and $z$ are shown as well.}
   \label{fig:sn13135_lc}
\end{figure}

  We note that the absolute $\chi^2$ values for the model fits do not appear
  to be very meaningful.  The models have not been calibrated against real
  data and there are no errors associated with the light curves (see, e.g.,
  Appendix~\ref{app:snia_lc}).  The current implementation also does not
  attempt to reject outlier points due to poor zeropointing and imperfect
  image registration, which are common features of the search photometry.  The
  best-fit $\chi^2$ value for the \snia\ model of a true \snia\ may be large
  compared to the number of degrees of freedom, but the confidence can be high
  if that value is well below the values of the other \sn\ types.  The {\it
    relative} values of the $\chi^2$ are, therefore, important and useful
  discriminators.  In Figure~\ref{fig:sn13135_chisq}, we show the values of
  $\chi_{\rm{min}}^2$, $\chi^2_1 < \chi^2_2$, and $0.5~(\chi^2_{\rm{min}} +
  \chi^2_1 + \chi^2_2)/3$ as functions of the number of epochs for \sn 2006fz,
  a Type Ia at z=0.105, with a flat galaxy photometric redshift prior.  For
  this candidate, the model with the minimum $\chi^2$ corresponds to the
  \snia\ model, and both the ``A'' and ``B'' criteria are satisfied at all
  epochs.  Also shown in Figure~\ref{fig:sn13135_lc} are the light curve fits
  to \sn 2006fz using the first 2, 4, 6, and 8 epochs of search photometry.

\begin{figure}[tb]  
  \begin{center}
    \includegraphics[angle=270, width=0.7\textwidth]{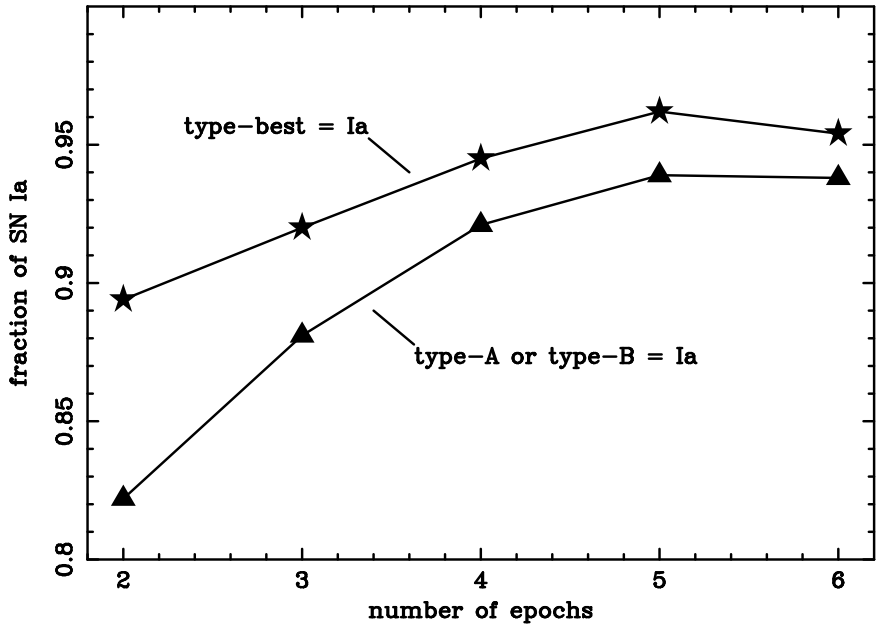}
  \end{center}
   \caption{The fraction of spectroscopically confirmed \sneia\ whose best-fit
     model is that of a \snia\ as a function of the number epochs (top curve
     with stars).  Also shown is the fraction of \sneia, which satisfy one of
     the ``A'' or ``B'' criteria with or without a galaxy redshift prior
     (bottom curve with triangles).  A large fraction of the sources ($>
     80$\%) are typed as \snia\ after only 2 epoch of imaging.}
   \label{fig:type_snia}
\end{figure}

\begin{figure}[tb]  
  \begin{center}
    \includegraphics[angle=270, width=0.55\textwidth]{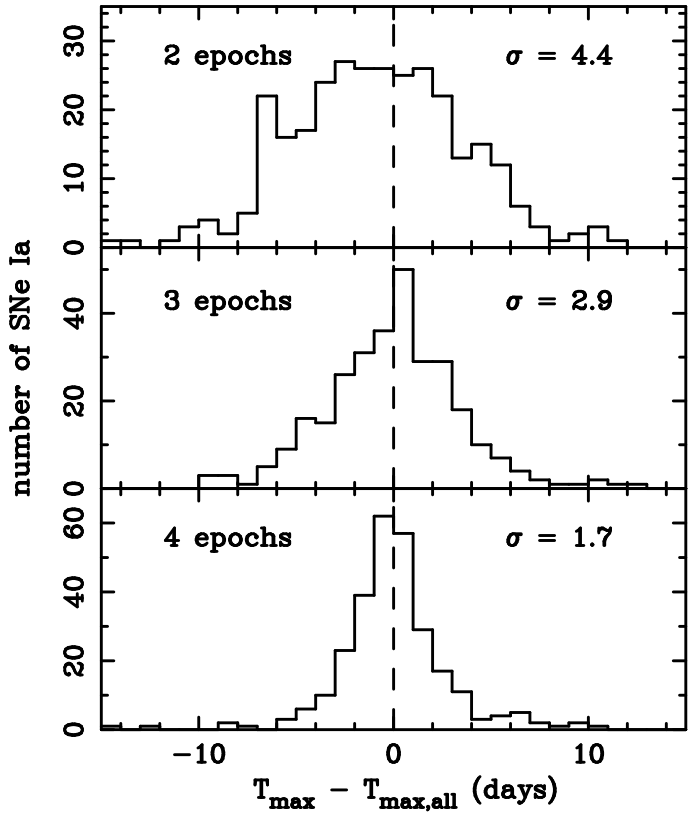}
  \end{center}
   \caption{Distribution of the difference between the time of $B$-band
     maximum estimated using the full light curves and that using the first 2
     (top), 3 (middle), and 4 (bottom) epochs of the spectroscopically
     confirmed \sneia.  Values are taken from fits with a flat redshift prior.
     The distributions are fit with a single gaussian, and the $\sigma$ values
     are listed in each panel.  The width of the distribution stays
     approximately constant above 4 epochs.  The vertical dashed lines
     represent the x-axis origin; not the centroid of the distributions.}
   \label{fig:peakmjd}
\end{figure}

  The overall performance of the photometric typing software is demonstrated
  in Figure~\ref{fig:type_snia}, where we plot the fraction of the
  spectroscopically confirmed \sneia, whose best-fit model is that of a \snia\
  as a function of the number of the light curve epochs.  The figure also
  shows the fraction of \snia\ that satisfy one of the ``A'' or ``B'' criteria
  with or without a nearest-neighbor host galaxy redshift prior.  Note that
  the fractions increase between 2 to $\sim 4$~epochs, but the improvement
  beyond $\sim 5$~epochs is marginal.  The fraction with the best type equal
  to a Ia does not reach unity because the spectroscopic sample includes 1)
  peculiar \sneia, 2) candidates with poorly subtracted images and photometry
  epochs, 3) \sneia\ discovered well after maximum light, and 4) \sne\ with
  low-\signoi\ light curves whose best-fit type drifts with the number of
  epochs.  Candidates that fall into the first three categories were targeted
  for spectroscopy because they were nearby and bright enough to be observed
  on 3-m class telescopes, and those candidates were observed even if their
  best-fit photometric type was not a \snia\ (see
  \S~\ref{subsec:spectro-strategy}).  The candidates in the fourth category
  were observed because they were at one time typed as a \snia.  We also show
  in Figure~\ref{fig:peakmjd} the time of $B$-band maximum light estimated
  from 2 -- 4 light curve epochs in comparison with the ``final'' estimate of
  the peak using the full light curves.  Again, the reliability improves
  substantially between 2 and $\sim 4$~epochs.

  Studies of the software and handscanning efficiencies as functions of the
  object brightness with the fakes (see Section~\ref{sec:fakes}) show that the
  \sn\ candidate identification and spectroscopic follow-up observations are
  essentially complete out to $z \sim 0.12$ \citep{dilday07}.  The one
  exception to this is our bias against highly-extincted \sne, since the grid
  of light curve models did not extend beyond $A_V = 1.0$ in 2005 and $A_V =
  3.0$ in 2006.  A post-season analysis of the 2005 candidates with an
  extended $A_V$ range recovered two additional photometric \snia\ candidates
  with large extinction in this redshift range.  There is no evidence for an
  additional population of low-$z$ \sneia\ that were missed due to the models
  adopted.

  The light curve fits generally run on the \gri\ search photometry, which is
  usually accurate to within $\sim 0.1$~mag.  Whenever the computer cluster at
  \apo\ is idle, we run the pipeline in {\tt sdssforce} mode (see,
  \S\ref{subsub:framesub}) to produce $u$ and $z$-band differenced images from
  all runs available on the mountain for each of the \sn\ candidates.  This
  typically includes frames from several pre-discovery epochs.  The $u$-band
  photometry is sometimes useful for distinguishing \sneia\ from core-collapse
  \sne\ that are generally bluer ($u - g \la 0.5$) during the early phases
  after explosion.

\subsection{Observability Index}
\label{subsec:observability}

  After the \sn\ photometric typing process, the software computes three
  quantities that help prioritize the target list for spectroscopic
  observations.  These quantities are referred to as ``weights'' and they are
  constructed specifically to identify candidates that are 1) near or before
  peak brightness, 2) less contaminated by galaxy light, and 3) not heavily
  extincted by intervening dust.  The product of the three weights defines the
  ``observability index''; candidates with higher observability index are
  generally assigned higher priority.

  For 1), we define the ``time weight'' as,
  \begin{equation}
    \label{equ:W_T}
    W_T = \left\{ \begin{array}{ll}
          e^{|t|\{\Delta m_{15}(B)/[10(1+z)]\}} & {\rm{if}}~t < 0~\rm{days}; \\
          e^{-t\{\Delta m_{15}(B)/[20(1+z)]\}}  & {\rm{if}}~t \geq 0~\rm{days},
    \end{array} \right .
  \end{equation}
  where $t$ is the current estimate of the \sn\ epoch in days relative to
  $B$-band maximum brightness.  Note that $W_T$ is large when the \sn\ is
  young, as \sn\ are most easily classified near maximum light
  \citep{filippenko97}, and decays exponentially with a characteristic time
  scale of 20 days.  The decline-rate parameter \dmB\ and redshift $z$ are
  also adopted from the best-fit values obtained from the light curve fits.
  If the best-fit type is a core-collapse \sn, \dmB\ is set to unity.

  The second weight computed by the software is the ``contamination weight''
  defined as,
  \begin{equation}
    \label{equ:W_C}
    W_C = e^{-2/\theta} \times
    \left(\frac{F_{{\rm{SN}},r}}{F_{{\rm{Gal}},r}}\right)^{1/2},
  \end{equation}
  where $\theta$ is the distance in arcseconds between the \sn\ and the
  centroid of the nearest-neighbor host galaxy.  The quantity
  $F_{{\rm{SN}},r}$ is proportional to the {\it current} $r$-band flux of the
  \sn, again, as estimated from the best-fit light curve model,
\begin{equation}
  \label{equ:F_SN}
  F_{{\rm{SN}},r} = 10^{-0.4~r_{\rm{SN}}}
\end{equation}
  and $F_{{\rm{Gal}},r}$ is the estimated local galaxy $r$-band flux at the
  position of the \sn\ defined to be,
\begin{equation}
  \label{equ:F_gal}
  F_{{\rm{Gal}},r} = 10^{-0.4(r_{\rm{Gal}} + d/d_{{\rm{eff}}})},
\end{equation}
  where $d$ is the angular separation between the \sn\ and the centroid of the
  nearest-neighbor source and $d_{\rm{eff}}$ is the effective radius of that
  source along position angle of the \sn.  This quantity is computed according
  to,
\begin{equation}
  \label{equ:d_eff}
  d_{\rm{eff}} = \frac{{\tt isoA}^2 (1-e^2)}{1-e^2\cos^2\phi}.
\end{equation}
  Here, $\phi$ is the position angle of the \sn\ measured from the position
  angle of host galaxy's semimajor axis ({\tt isoPhi}) with ellipticity given
  by,
\begin{equation}
  \label{equ:ellip}
  e^2 = 1 - \left(\frac{{\tt isoB}}{{\tt isoA}}\right)^2.
\end{equation}
  The isophotal galaxy parameters ({\tt isoA}, {\tt isoB}, and {\tt isoPhi}),
  which are used to measure the galaxy's ellipticity of the 25 magnitudes per
  arcsecond isophote, are adopted from the \sdss\ {\small DR}4
  catalog\footnote{http://www.sdss.org/dr4/algorithms/classify.html}.
  Clearly, candidates with larger values of $W_C$ suffer less contamination
  from the host galaxy light.

  The third and final weight is the ``dust weight'', which estimates the
  amount of dust extinction from both host galaxy morphology and color, and is
  defined to be,
  \begin{equation}
    \label{equ:W_D}
    W_D = e^{-d_{\rm{eff}}/d} \times (1 + 3~f_{\rm{deV}} + 5~c_{\rm{ellip}}),
  \end{equation}
  The quantity $f_{\rm{deV}}$ is the fractional probability of the source
  being well represented by a de Vaucouleurs profile and is again computed
  from values in the \sdss\ catalog, i.e.,
  \begin{equation}
    \label{equ:f_deV}
    f_{\rm{deV}} = \frac{{\tt deV\_L}}{{\tt deV\_L} + {\tt exp\_L} + {\tt
    star\_L}},
  \end{equation}
  where {\tt deV\_L}, {\tt exp\_L}, {\tt star\_L} are the null hypothesis
  probabilities of the object being well-represented by a de Vaucouleurs
  profile, an exponential function, and a stellar profile (\psf),
  respectively.  The last quantity in parentheses of Equation~\ref{equ:W_D}
  characterizes the color of the galaxy, and is defined to be,
  \begin{equation}
    c_{\rm{ellip}} = \left\{ \begin{array}{ll}
      (r_{\rm{Gal}} - i_{\rm{Gal}} - 0.4)/0.2 & {\rm{if}}~(r_{\rm{Gal}} -
      i_{\rm{Gal}}) > 0.4~{\rm{and}}~(g_{\rm{Gal}} - r_{\rm{Gal}}) > 0.9; \\
      0 & {\rm{otherwise}},
    \end{array} \right .
  \end{equation}
  where $g_{\rm{Gal}},r_{\rm{Gal}},i_{\rm{Gal}}$ are the host galaxy $g$, $r$,
  and $i$-band model magnitudes.  This quantity is non-zero only for red
  elliptical galaxy candidates, which typically populate the color-color space
  bounded by $(r - i) > 0.4$ and $(g - r) > 0.9$.  Note that the light curves
  also provide an independent (and usually more reliable) estimate of the
  amount of dust extinction.

  The final observability index is defined as the product of the three weights,
  \begin{equation}
    {\rm{observability~index}} = W_T \times W_C \times W_D.
  \end{equation}
  In practice, the observability index is rarely used for the bright ($r \la
  20.5$~mag) low-$z$ \sn\ candidates, since there are enough spectroscopic
  resources to observe nearly all of those targets.  At high $z$, however, the
  number of \snia\ candidates exceeds the number we can observe given our
  limited resources on large-aperture telescopes.  In order to select the best
  targets from a large number of candidates that satisfy the light curve
  selection criteria, the sources are ranked based on the value of the
  observability index.  Manual screening and selection of the sources by the
  spectroscopic coordinator is still required.  We describe our strategy in
  detail below.

\subsection{Spectroscopic Follow-up Strategy}
\label{subsec:spectro-strategy}

  Spectra are obtained by the various spectroscopic teams, who independently
  apply for observing time during each of the search seasons.  The teams
  agreed to confirm types and obtain redshifts of \snia\ candidates at the
  highest priority, but each team generally has one or more of their own
  \sn\ projects that they pursue in parallel.  Some of these include 1)
  multi-epoch spectroscopy of peculiar \sne, 2) \sne\ in underluminous host
  galaxies, 3) \snibc\ and hypernova candidates, 4) detailed spectroscopic
  properties of \snia\ hosts, and 5) multi-epoch studies of line features and
  their diversity in \sneia.  Whenever possible, we also try to obtain spectra
  of other variable objects if a scanner has classified them as
  ``interesting''.  An example of this is a dwarf nova discovered in Aquarius
  \citep{prieto06}.

  In 2005, the primary telescopes used for observing the low-redshift
  candidates were the \arc\ 3.5m, \wht\ 4.2m and the \mdm\ 2.4m telescopes.
  The high-z candidates were observed with either the \het\ or Subaru.  The
  resources were sufficient to obtain spectra of most $z \la 0.15$ \snia\
  candidates as well as core-collapse \sne\ at $z \la 0.06$, but only a
  fraction of the $z \ga 0.15$ candidates were observed.

  During the Fall 2006 campaign, the various spectroscopic teams were awarded
  sufficient 3m-class telescope time to obtain a spectrum of essentially {\it
    all} \sn\ candidates with $r \la 20.5$~mag (an average \snia\ at $z \la
  0.15$ and core-collapse \sn\ at $z \la 0.06$ near maximum light) {\it and}
  obtain multi-epoch spectra of a large sample of nearby ($z \la 0.1$) \sneia.

  We generally require at least two epochs of detection before we target the
  candidate for spectroscopic observation, to guarantee that the source is not
  an asteroid.  This also enables a better estimate of the epoch of peak
  brightness of the \sn.  Since the primary goal of our survey is to obtain
  well-sampled light curves of Type Ia \sne, \sn\ Ia candidates that are
  discovered early are generally given higher priority for spectroscopy.  This
  is especially true for the very low-$z$ candidates ($z \la 0.1$) to enable
  rapid confirmation and dissemination to the public.

  In practice, we initially select \snia\ candidates that satisfy at least one
  of the ``A'' or ``B'' criteria either with or without a galaxy photometric
  redshift prior.  There are times, however, during stretches of poor weather
  at \apo\ and/or when several 3m-class spectroscopic telescopes are online,
  when there are not enough new and bright \snia\ candidates to observe.  This
  was not an issue during most of the 2005 season, but it occurred frequently
  in 2006.  In such cases, we loosen the selection criteria and inspect all
  currently-bright sources irrespective of their best-fit \sn\ type; there is
  only small number ($\sim 30 - 50$) of sources with $r \la 20.5$~mag at any
  given time.  When necessary, we also target candidates that have only a
  single epoch of detection.  Extremely red single-epoch targets that are not
  associated with a host galaxy are generally avoided (to minimize the chance
  of observing an asteroid).

  On the other hand, there is never a shortage of higher-$z$ \snia\ targets at
  $z \ga 0.2$, so the ``A'' or ``B'' criterion with or without a galaxy
  photometric redshift prior is always used in the pre-selection process.  At
  these redshifts, we generally avoid candidates that are located within $\sim
  0.5$\arcsec\ of the centroid of a host galaxy, since they have a relatively
  higher chance of being an active galactic nucleus.  Candidates at $z \sim
  0.3$ that are well-separated from the host and suffer a relatively low level
  of extinction (as estimated from the multi-band light curve fits) are
  high-priority (priority 0 and 1) targets for \het, which is a
  queue-scheduled telescope \citep{shetrone07}.  Lower-$z$ candidates ($z \sim
  0.2$) are assigned lower \het\ priority (mostly priority 2, but some at
  priority 3).  Targets are generally kept active in the \het\ queue until
  they are $\sim 20$ days past maximum light as estimated from the light curve
  fits.  The Subaru telescope has focused on \snia\ candidates in a similar
  redshift range, but, to take advantage of its superb image quality, the
  Subaru telescope has generally been assigned targets that suffer a
  relatively large amount of host contamination (i.e., a low value of $W_C$;
  see Section~\ref{subsec:observability}).

  The spectra are often analyzed in real time by the observers on site, and
  feedback is given to the spectroscopic coordinator.  Depending on the
  observing conditions, the target list is adjusted in real time.  At the end
  of the night, the observers provide a preliminary spectroscopic type of each
  of the candidates observed.  The type and the preliminary redshift is
  entered immediately into the database, and the results are disseminated to
  the community through the Central Bureau Electronic Telegrams.  All nearby
  \sne\ ($z \la 0.08$), however, are also announced to a list of subscribers
  via email usually within hours from confirmation, allowing rapid
  complementary follow up on other telescopes.  Information on all
  spectroscopically confirmed \sne\ is placed on a public web
  page\footnote{http://sdssdp47.fnal.gov/sdsssn/snlist\_confirmed.php} as soon
  as the \sn\ type has been entered into the database.  A list of all
  spectroscopically confirmed \sne\ from 2005 and 2006 is presented in
  Tables~\ref{tbl:sn2005} and \ref{tbl:sn2006}, respectively.

\section{Imaging Follow-up}
\label{sec:imaging}

  Throughout the 2005 search, we used some of our \arc\ 3.5m and \mdm\ 2.4m
  time, as well as the imager on the {\small NMSU} 1m telescope, University of
  Hawaii 88-inch telescope, the 1.8m Vatican Advanced Technology Telescope
  (\vatt), and the 3.5m \wiyn\ telescope to (1) augment the light curve points
  of spectroscopically confirmed \sneia\ during periods of poor weather
  conditions at \apo\ and (2) obtain late-time photometry of \sne\ that had
  faded below the \sdss\ detection limit.  Some imaging observations were
  performed by the \vatt, the \wiyn, and the 1.5m telescope at Maidanak
  Observatory during the 2006 season.  Additional imaging data were obtained
  on the \arc\ 3.5m during November of 2006 when the Dual Imaging Spectrograph
  was not functional.  This effort of follow-up imaging was not carried over
  for the \mdm\ and for \arc\ during September and October, since we concluded
  that spectroscopy is the more valuable use of our resources.  In December of
  both 2005 and 2006, however, imaging observations were performed to get
  light curves of \sne\ discovered in November past maximum light.  Deep
  multi-band imaging of several host galaxies of \sneia\ were also obtained by
  the 2.5m Isaac Newton Telescope.

  A few high-profile \sdssii\ \sn\ targets were also observed extensively by
  several follow-up programs -- optical imaging and spectroscopy by the CfA
  \sn\ Group, optical and infrared imaging and optical spectroscopy by the
  \csp, and optical spectroscopy by the \snf.

\section{Results from Fall 2005 and 2006}
\label{sec:results}

  The 2005 season resulted in 73 unique \sdss\ imaging runs acquired on 59
  different nights.  Approximately half of the frames were taken under
  non-photometric conditions, bright moon, and/or poor seeing.  A total of
  155,616 objects were visually inspected during this season, of which 24,402
  were tagged as potential \sne, resulting in 11,385 unique \sn\ candidates
  (see Table~\ref{tbl:summary}).  Interestingly, 6,618 of those candidates
  were detected in only a single epoch, and are mostly likely moving solar
  system objects that were not filtered properly.  A few of these single-epoch
  candidates may be fast transients that rise and fade on timescales shorter
  than $\sim 1$~day \citep{becker04}.  Most of the remaining 4767 candidates
  are true variables or transient sources\footnote{A very small fraction of
  the 4767 multi-epoch candidates are artefacts of image subtraction -- e.g.,
  faint dipoles that appear in the same place at two or more different
  epochs.}.  A short summary of the 2005 and 2006 seasons is shown in
  Table~\ref{tbl:summary}.

  For the 2005 search season, we acquired 248 spectra from 187 distinct \sn\
  candidates under various observing conditions.  A total of 130 unique
  candidates were spectroscopically confirmed to be of Type Ia.  One of these
  \sneia\ (\sn 2005hj) was co-discovered and spectroscopically observed by the
  Texas Supernova Search \citep{quimby07}.  The number of confirmed
  core-collapse \sne\ were 7 and 13 for Type Ib/c and II, respectively.  One
  of these Type II \sne\ (\sn 2005mj) was spectroscopically confirmed by the
  \essence\ Group.  An additional 40 candidates were observed in the
  post-season during December 2005 -- January 2006.  A sample of 41 \sn\
  targets including \sn 2005hk \citep{phillips07} and \sn 2005gj
  \citep{prieto07} for which we acquired 16 and 9 spectra, respectively, were
  observed spectroscopically more than once.  The host galaxy of one 2005
  candidate (\sn 7017) was observed during the 2006 search season.
  Interestingly, that supernova was still active and visible a year after
  discovery; the spectrum was identified as a peculiar \sn\ similar to
  \sn2002ic and \sn2005gj.

  In 2006, 90 \sdss\ imaging runs were taken on 60 nights.  The number of
  visually-inspected objects was reduced by more than an order of magnitude to
  14,430.  This improvement is due to the modified moving object finder and
  the choice of scanning only second-epoch and bright first-epoch objects.  We
  identified a total of 3694 unique \sn\ candidates.  Surprisingly, there is
  still a somewhat large sample (599 out of 3694) of slow-moving single-epoch
  objects that were tagged by a scanner as a possible \sn\ and made the
  candidates list.

  During the second search season, a total of 449 spectra were acquired, $\sim
  80$\% more than the 2005 season, from 285 unique candidates.  Though the
  majority of these were observations of new 2006 candidates, a handful of
  host galaxies of the 2005 candidates were observed as well.  A sample of 197
  candidates, including five candidates that were observed by the CfA \sn\
  Group and the \essence\ Survey, were spectroscopically confirmed to be
  \sneia\ and 14 candidates were identified as spectroscopically probable
  \sneia.  The number of confirmed core-collapse \sne\ were 7 and 19 for Type
  Ib/c and II, respectively.

  Table~\ref{tbl:spectro} summarizes the spectroscopic observations grouped
  into candidates discovered in 2005 and 2006.  We performed 172 (241)
  observations of new \snia\ candidates discovered in 2005 (2006), of which
  130 (197) were spectroscopically confirmed to be \sneia, and 16 (14)
  candidates were identified as spectroscopically probable \sneia; the latter
  are denoted \sn\ Ia? in Tables~\ref{tbl:sn2005} and \ref{tbl:sn2006}.  In
  addition, observations of 7 (7) additional candidates in 2005 (2006)
  resulted in galaxy spectra (either the \sn\ had faded or the galaxy was too
  bright relative to the \sn) with spectroscopic redshifts consistent with
  photometric redshifts estimated from the multi-band light curves of the
  \sne.  Therefore, in the first two years, 90\% of \snia\ targets resulted in
  spectra of \sneia, probable \sneia, or host galaxies of photometric \snia\
  candidates.  A total of 12 \snia\ targets (3\%) were classified instead as
  core-collapse \sne, 4 targets (1\%) were classified as either an \agn\ or a
  flaring M-dwarf, and 9 targets (2\%) resulted in galaxy spectra with
  spectroscopic redshifts inconsistent with photometric redshifts estimated
  from the \snia\ model fits.  Finally, there is a sample of 18 spectra (4\%),
  most of which were taken under poor weather conditions, that are
  unidentified.

  Evaluation of the full light curves after the end of each search season
  enables identification of additional photometric \snia\ candidates that, for
  various reasons, were not spectroscopically observed during the search.  We
  have identified at least $\sim 200$ high-quality \snia\ candidates from the
  2005 candidates list, and have already acquired host galaxy spectra for a
  significantly-sized subsample.  These targets are also good fillers during
  periods of poor observing conditions at \apo\ that result in a lack of good
  new \sn\ candidates.  As of the writing of this paper, we have measured
  redshifts for 81 host galaxies of candidates with Ia-like light curves.  An
  additional 13 candidates were identified to be good Ia candidates with host
  galaxy redshifts from the \sdss\ redshift survey.

\section{Summary}
\label{sec:summary}

  The search pipeline of the \sdssii\ \sn\ Survey has enabled efficient
  discoveries of variable and transient astronomical sources, filtering over
  375,000 objects detected each season into several thousands of \sn\
  candidates.  A reliable photometric \sn\ typing system and spectroscopic
  follow-up algorithm have allowed spectroscopic observations of $\sim 150$
  \sneia\ per season with 90\% targeting efficiency.  After two seasons, the
  search and spectroscopic follow-up algorithms have reached a relatively
  mature stage, and it is unlikely that major changes will be made for the
  third and final season of 2007.  If the observing conditions resemble those
  of the previous two years, we expect to increase our sample of
  spectroscopically confirmed \snia\ by an additional $\sim 150 - 200$ events,
  reaching a sample of $\sim 500$ confirmed \sneia\ for the completed survey.
  A majority of the low-redshift sources will have well-sampled multi-band
  light curves that can serve as templates for future studies.

  Spectroscopic target selection by a human is appropriate (and preferred) for
  a relatively small survey like the \sdssii\ \sn\ survey, but this is
  unlikely to be feasible for future large-scale surveys that will discover
  thousands or tens of thousands of \sne\ over the period of a few years.
  Based on our experience, however, we believe that much of the candidate
  identification process can be automated, and with just 2 -- 4 epochs of
  multi-band imaging, \sn\ candidates can be assigned probabilities that are
  reliable enough for performing follow-up spectroscopy.  More quantitative
  studies of \sn\ identification using photometric data alone will be
  presented in a future article.

  \clearpage

  \acknowledgements Funding for the \sdss\ and \sdssii\ has been provided by
  the Alfred P. Sloan Foundation, the Participating Institutions, the National
  Science Foundation, the U.S. Department of Energy, the National Aeronautics
  and Space Administration, the Japanese Monbukagakusho, the Max Planck
  Society, and the Higher Education Funding Council for England. The
  \sdss\ Web Site is \verb9http://www.sdss.org/9.

  The \sdss\ is managed by the Astrophysical Research Consortium for the
  Participating Institutions. The Participating Institutions are the American
  Museum of Natural History, Astrophysical Institute Potsdam, University of
  Basel, Cambridge University, Case Western Reserve University, University of
  Chicago, Drexel University, Fermilab, the Institute for Advanced Study, the
  Japan Participation Group, Johns Hopkins University, the Joint Institute for
  Nuclear Astrophysics, the Kavli Institute for Particle Astrophysics and
  Cosmology, the Korean Scientist Group, the Chinese Academy of Sciences
  ({\small LAMOST}), Los Alamos National Laboratory, the Max-Planck-Institute
  for Astronomy ({\small MPIA}), the Max-Planck-Institute for Astrophysics
  ({\small MPA}), New Mexico State University, Ohio State University,
  University of Pittsburgh, University of Portsmouth, Princeton University,
  the United States Naval Observatory, and the University of Washington.

  The Hobby-Eberly Telescope (\het) is a joint project of the University of
  Texas at Austin, the Pennsylvania State University, Stanford University,
  Ludwig-Maximillians-Universit\"at M\"unchen, and Georg-August-Universit\"at
  G\"ottingen.  The \het\ is named in honor of its principal benefactors,
  William P. Hobby and Robert E. Eberly.  The Marcario Low-Resolution
  Spectrograph is named for Mike Marcario of High Lonesome Optics, who
  fabricated several optics for the instrument but died before its completion;
  it is a joint project of the Hobby-Eberly Telescope partnership and the
  Instituto de Astronom\'{\i}a de la Universidad Nacional Aut\'onoma de
  M\'exico.  The Apache Point Observatory 3.5-meter telescope is owned and
  operated by the Astrophysical Research Consortium.  We thank the observatory
  director, Suzanne Hawley, and site manager, Bruce Gillespie, for their
  support of this project.  The Subaru Telescope is operated by the National
  Astronomical Observatory of Japan.  The William Herschel Telescope is
  operated by the Isaac Newton Group, and the Nordic Optical Telescope is
  operated jointly by Denmark, Finland, Iceland, Norway, and Sweden, both on
  the island of La Palma in the Spanish Observatorio del Roque de los
  Muchachos of the Instituto de Astrofisica de Canarias.  Observations at the
  {\small ESO} New Technology Telescope at La Silla Observatory were made
  under programme {\small ID}s 77.A-0437, 78.A-0325, and 79.A-0715.  Kitt Peak
  National Observatory, National Optical Astronomy Observatory, is operated by
  the Association of Universities for Research in Astronomy, Inc. ({\small
  AURA}) under cooperative agreement with the National Science Foundation.
  The {\small WIYN} Observatory is a joint facility of the University of
  Wisconsin-Madison, Indiana University, Yale University, and the National
  Optical Astronomy Observatories.  The W.M.\ Keck Observatory is operated as
  a scientific partnership among the California Institute of Technology, the
  University of California, and the National Aeronautics and Space
  Administration.  The Observatory was made possible by the generous financial
  support of the W.M.\ Keck Foundation.  The South African Large Telescope of
  the South African Astronomical Observatory is operated by a partnership
  between the National Research Foundation of South Africa, Nicolaus
  Copernicus Astronomical Center of the Polish Academy of Sciences, the
  Hobby-Eberly Telescope Board, Rutgers University, Georg-August-Universit\"at
  G\"ottingen, University of Wisconsin-Madison, University of Canterbury,
  University of North Carolina-Chapel Hill, Dartmough College, Carnegie Mellon
  University, and the United Kingdom \salt\ consortium.

\clearpage

\appendix
\section{Supernova Software}
\label{app:software}

  This appendix describes the details of the supernova software that runs on
  the computer cluster at \apo.

  The majority of {\tt Framesub} is written in the {\tt Perl} language.  This
  provides the internal glue that strings together the various processing
  steps.  In general, the image-level computations are written in the {\tt C}
  language.  These applications are called by the {\tt Perl} scripts.  The
  implementation of {\tt Framesub} on the computing cluster is controlled by
  shell scripts.

  As a programmatic summary, the {\tt Framesub} pipeline consists of a series
  of {\it stages}, each of which has {\it actions} which it undertakes, as
  well as {\it dependencies} on the successful completion of previous stages.
  By default, an ensemble of images is passed from stage to stage using input
  and output lists.  If an image fails the assigned actions in a particular
  stage, the frame is {\it not} passed into the input list for the subsequent
  stages of processing.  A record is kept of the fact that it failed, and all
  such failures must be addressed individually.  Below we describe the three
  modes in which the pipeline is run.

\subsection{{\tt sdssred}}
\label{sec:apsdssred}

  This mode performs the basic reduction and preparation of the data for
  image subtraction.

\begin{itemize}

\item{{\tt FINDNEWIM}:} After processing by \photo, {\tt Framesub} must
  initialize its input list of images.  This stage searches for the correct
  data products\footnote{http://www.sdss.org/dr6/dm/flatFiles/FILES.html} for
  the camera column--filter combination assigned to it.  This includes the
  {\tt fpC} science image, a {\tt fpM} mask image, an {\tt asTrans} file
  providing the astrometry, and a {\tt psField} file that contains \photo 's
  model of the \psf\ \citep{stoughton02}.

\item{{\tt MKSATMASK}:} The {\tt fpM} file contains a pixel--by--pixel mask
  corresponding to any operations that have happened on the science image.
  This includes interpolation over \ccd\ defects and cosmic rays, the masking
  of saturated pixels, and interpolation over bleed trails from the saturated
  stars.  This stage translates the \sdss --format mask file into a mask image
  understood by {\tt Framesub}.  In practice, this is merely a reassignment to
  the numerical values of the mask bits.

  We provide an additional level of editing in these images by masking halos
  around clusters of saturated pixels, as well as extending diffraction spike
  masks at the orientation correct for the camera rotation.  These additional
  masks are only added for clusters of more than 20 saturated pixels.  An
  operational scale length $r$ for the halo and diffraction masking is
  determined by using the number of saturated pixels $N_{\rm sat}$ and solving
  for $r$ through $N_{\rm sat} = \pi r^2$.

\item{{\tt SDSSPSF}:} This stage converts the solution for the \psf~obtained
  by \photo\, contained in the {\tt psField} files, into a format readable by
  {\tt Dophot} \citep{schechter93}.  The \photo\ \psf~solution is later used
  to make an aperture correction to the difference image photometric
  measurements (see section \ref{sec:apdiffim}).

\item{{\tt SDSSSTARS}:} This stage obtains a list of the positions,
  magnitudes, and magnitude errors for the calibration stars contained within
  the frame and stores them in a text file suitable as input to {\tt
  SDSSZERO}, which computes the zeropoints of each of the frames (see
  below). The list of calibration stars is also an optional input to {\tt
  DIFFIM}, providing a list of preferred sources to be used in deriving the
  optimal convolution kernel (see section \ref{sec:apdiffim}). The magnitudes
  and magnitude errors were initially based on the {\small DR}4 database
  \citep{dr4}, but were later updated based on the catalog provided by
  \citet{ivezic07}.

  The difference images obtained by the \sn\ Survey are always photometrically
  scaled to the template image; the following two stages are included when
  running {\tt sdssred} on the template images.

\item{{\tt STARDOPHOT}:} This stage performs doPhot photometry on a list of
   calibration stars provided by {\tt SDSSSTARS}, using the modified version
   of the {\tt Dophot} software package.

\item{{\tt SDSSZERO}:} This stage performs zeropointing of the frame.  The
  magnitudes and magnitude errors of the calibration stars are first converted
  to flux units via $10^{-0.4 m}$, and then the model, ADU counts = A*flux, is
  solved for A using the weighted least-squares procedure. The zeropointing
  algorithm does iterative sigma clipping at $\{24.0,12.0,6.0,6.0,6.0\}$
  standard deviations from the best fit zeropoint model.  The final zeropoint,
  the statistical error on the zeropoint, and the RMS of the zeropoint are
  then converted into magnitudes and stored for use during the \sn\ search.

\end{itemize}

\subsection{{\tt sdssred} With Fakes}
\label{sec:fakes}

  To monitor the software and human search efficiencies, we insert artificial
  \sne\ (fakes) into the data stream in real-time.  Prior to the beginning of
  the survey, a library of fakes was generated (1000 fakes in 2005).  Each
  fake is assigned a redshift, celestial position ($\alpha$, $\delta$), date
  of peak-magnitude in $V$ band, and an intrinsic luminosity.  The redshift
  distribution was generated according to $dN/dz \propto z^2$ and the
  celestial positions are then chosen to be near to a galaxy with the same
  redshift, based on a catalog of photometric redshifts \citep{oyaizu07}.

  The fakes generated during the 2005 observing campaign did not include
  extinction and reddening; however the set of fakes generated during the 2006
  campaign included extinction, with $A_V$ chosen from an exponential
  distribution, $P(A_V) \propto e^{-A_V/0.4}$.

  The following stages are used to generate realistic Type Ia \sn\ light
  curves for the fakes in real time.

\begin{itemize}

\item{{\tt SDSSFAKESELECT}:} To minimize computation time for inserting fakes
  into the data stream and still retain the ability to simulate the magnitude
  at an arbitrary light-curve epoch, a range of dates for which the fake
  magnitude is within the expected limit of the \sdss\ telescope was computed.
  This stage checks the library of fakes to see whether any fake has a
  position within the limits of the field and that the present \mjd\ falls
  within the expected range of observability.

\item{{\tt STARDOPHOT\_FAKE}:} This stage performs {\tt Dophot} photometry on
  calibration stars within the frame. This is necessary to dynamically derive
  the magnitude to \adu\ conversion for the fakes.

\item{{\tt SDSSZERO\_FAKE}:} This stage computes a zeropoint for the frame,
  based on the dophot photometry from the previous stage.  It is identical to
  the {\tt SDSSZERO} stage described in section \ref{sec:apsdssred}.

\item{{\tt SDSSFAKEINSERT}:} This stage inserts the fake \sne\ into the search
  image.  Given the current \mjd\ and the fake light-curve parameters
  described above, the theoretical magnitude is computed in real-time. The
  light-curve model is identical to the stretch model used for online
  photometric typing, as described in \S~\ref{subsec:snphoto}.  Given the
  magnitude and the zeropoint computed above, an integrated flux value is
  computed.  The \psf\ model for the fake comes from the {\tt Photo} derived
  \psf; this is remapped to the astrometric grid of the search image and
  photometrically scaled.  Poisson noise is added to each pixel.  Finally, the
  unperturbed search image is copied and the \adu\ values for each pixel are
  overlaid on the search image, which is passed to the subsequent processing
  stages.  A record of changes to the image is written to the {\tt fpC} file
  header, and the magnitudes and corresponding \mjd\ are written to a {\tt
  MySQL} database.

\end{itemize}

\subsection{{\tt sdssdiff}}
\label{sec:apdiffim}

  This operational mode works with two sets of images that have been
  successfully run through the {\tt sdssred} stages.  One set of data is
  considered the ``template'' for difference imaging, and they are subtracted
  from the science images to yield only those objects that have varied in
  position or brightness.

\begin{figure}[t]  
  \begin{center}
    \includegraphics[angle=0, width=1\textwidth]{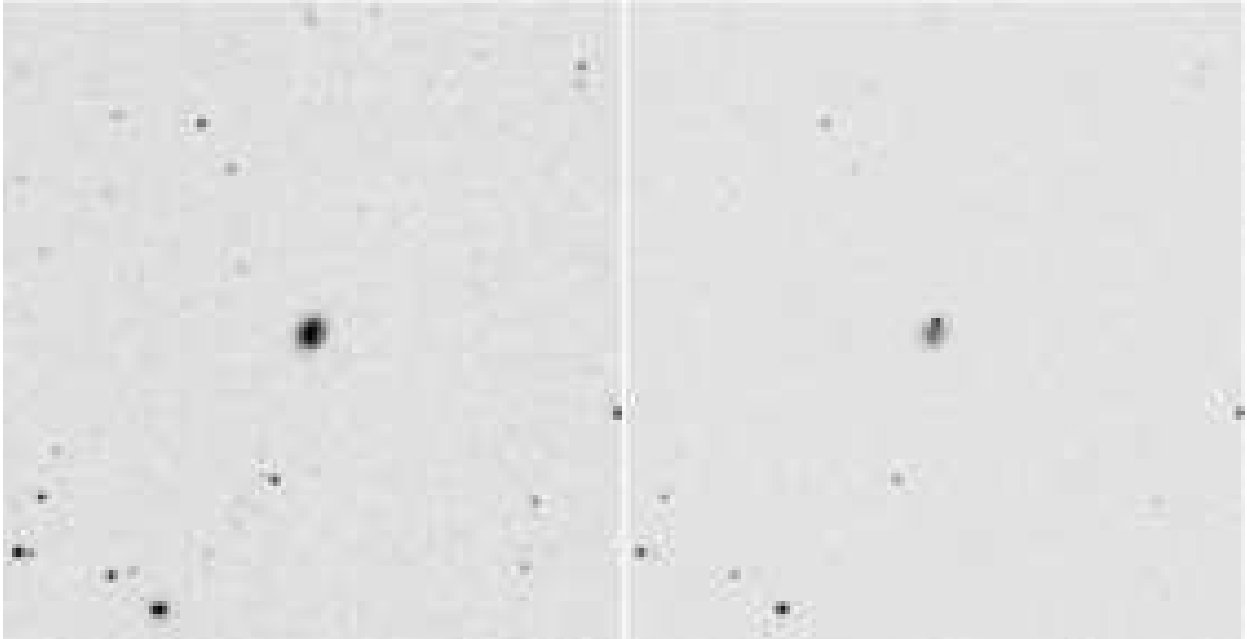}
  \end{center}
  \caption{Template (left) and search (right) $r$-band images centered on the
    host galaxy of \sn2005ir, a Type Ia \sn\ at $z = 0.076$.  The \sn\ is
    visible above and slightly to the right of the of the host galaxy in the
    right-hand image.  The search image was part of run 5823, and was taken
    under photometric conditions with $\sim 1.1$\arcsec\ seeing.  Images are
    3.3\arcmin\ on the side.  Note the enhanced density of faint sources ($r
    \ga 23$~mag) visible in the template image that are not detected in the
    search image.}
  \label{fig:sn7876comp}
\end{figure}

  In practice, most of our template images consist of the \psf--optimized
  coadds of \sdss\ Stripe 82 \citep{annis06}.  The images consist of 5 -- 10
  coadded exposures, depending on the position in the stripe, from runs prior
  to the 2005 \sn\ search season.  Only images with seeing $\la 1.1$\arcsec\
  are used.  Figure~\ref{fig:sn7876comp} shows an $r$-band coadd image of the
  field around \sn2005ir in comparison with a single-exposure search image.
  Some of our fields near the edge of the stripe fall outside of the coadd,
  and for these we use single--epoch templates from a night with good seeing.

\begin{itemize}

\item{{\tt MATCHTEMPL}:} Since the \sdss\ camera is used in drift-scan mode,
  the data from a given run comprises a long swath of the sky.  This tapestry
  is split by \photo\ into individual fields of $2048 \times 1489$ pixels,
  with 128 pixels of overlap between neighboring fields.  Because Stripe 82 is
  an equatorial field, this corresponds to a slice in $\alpha$.  It is not
  guaranteed that the $\alpha$ boundaries of the science image agree exactly
  with those of the template, meaning in general we will need to identify and
  combine the two template images that overlap the science image.  However,
  the pointing model of the telescope is sufficiently accurate that the
  overlap in $\delta$ is nearly complete.

\item{{\tt GLUETEMPL}:} This stage implements the pixel--level coaddition of
  the overlapping template images into a single template image whose \ra\
  coverage is similar to the input science image.

\item{{\tt REMAP}:} Because the images may still be misaligned in \ra\ by a
  fractional pixel offset, and in declination by multiple pixels, we apply a
  remapping kernel that {\it exactly} aligns the glued template image with the
  science image.  For this operation, we use the astrometry included in the
  {\tt asTrans} files to determine, for each pixel in the science image, what
  position in the template image needs to be resampled.  We apply a
  windowed--sinc resampling of the template image with a $9 \times 9$-pixel
  footprint (i.e., a Lanczos 4 kernel) to yield a final remapped glued
  template image.  The template is always remapped to the reference frame of
  the science image, to avoid convolving the science data.  Because the
  template image generally has much higher signal--to--noise (except possibly
  near the beginning of the stripe where only single-exposure templates are
  available), the ``smoothing'' effects due to convolution are not a serious
  issue.  We propagate the pixel noise and pixel masks through this
  convolution process.

\item{{\tt DIFFIM}:} After the {\it astrometric} registration described above,
  we implement a {\it photometric} registration that matches the \psf\ of the
  two images, so that they may be subtracted on a pixel--by--pixel basis.
  This uses a modified version of the \cite{alard00} algorithm as described in
  \cite{becker04}.

  In this procedure, known objects (stars or galaxies) are used to determine a
  set of convolution kernels that minimize the difference between each
  convolved template object and the respective science object.  Objects that
  include masked pixels are rejected, as are variable objects.  The ensemble
  of kernels is then used to construct a single kernel function that varies
  spatially with position on the image, to account for differences in the
  \psf\ across the frames.  This function is allowed to vary quadratically
  across the image.  We then apply this kernel to each pixel in the image and
  subtract the convolved template from the science image, yielding a
  difference image in which the static signal has been subtracted, and only
  objects that have varied in position or brightness remain.  We choose to
  always apply this kernel to the template image, even though in good seeing
  conditions this requires us to actually {\it de}convolve the template data.
  We propagate the image noise and mask similar to the {\tt REMAP} process.

\item{{\tt DIFFDOPHOT}:} We next perform object detection on the difference
  image, again using the modified {\tt Dophot} package.  The software is able
  to read the propagated noise and mask images to assist in the object
  detection and measurement.  The background is explicitly set to zero,
  assuming that the difference imaging has produced a background--subtracted
  image.  The \photo\ \psf\ model and aperture corrections are used to perform
  the photometry.  We only perform photometry on positive--going excursions,
  thus objects that fade in brightness are not measured.

\item{{\tt PIXCHK}:} We next read in the difference, noise, and mask images
  surrounding each detected object and examine the neighboring pixels.  The
  total {\it number} of positive--valued pixels, negative--valued pixels, and
  masked pixels are calculated in an aperture, as well as the total {\it
  amount} of flux in positive--valued and negative--valued pixels.  These will
  be used to reject dipoles due to image misalignment or general failures in
  the difference imaging.

\item{{\tt DIFFCUT}:} This stage rejects candidates from the photometry list
  based upon photometry flags and {\tt PIXCHK} values.  In particular, objects
  for which {\tt Dophot} is unable to perform a 4--parameter fit using the
  mean \psf\ parameters are rejected ({\tt Dophot} types 4 or 6).  Objects
  with \signoi\ smaller than 3.0 are also rejected, as well as objects with
  excessive fractions of masked pixels (0.6) or fractions of flux in negative
  pixels (0.65), amongst other tests.  These cuts yield the set of objects
  ingested into the database by {\tt DoObjects}.
\end{itemize}

\subsection{{\tt sdssforce}}

  This mode is used to perform forced--positional \psf\ photometry.  The
  coordinates of the candidate are determined from a \signoi--weighted average
  of all objects associated with the \sn.  The \psf\ model is adopted from
  output of \photo.  In practice, this mode is used to measure magnitudes in
  the $u$ and $z$ bands, which are not automatically processed by {\tt
  sdssdiff}, as well as $ugriz$ magnitudes in images obtained before the \sn\
  discovery to obtain measurements of low-\signoi\ detections and upper
  limits.  This mode yields more accurate measurements of the fluxes than
  those produced by {\tt sdssdiff}, and they generally agree with the final
  photometry to within $\sim 0.05$~mag in all filters \citep{holtzman07}.

\section{Autoscanner}
\label{app:autoscanner}

  As a further illustration of the histogram method of \pdf\ estimation, we
  note that Bayes' theorem, applied to classification, says that,
\begin{equation}
\frac{P(c_i|\theta)}{P(c_j|\theta)} =
\frac{P(\theta|c_i) P(c_i)}{P(\theta|c_j) P(c_j)}
\end{equation}
  where $P(a|b)$ is the conditional probability for $a$ given $b$, $c_i$
  represents object class $i$, and $\theta$ denote particular values for the
  set of observables of an object.  If this ratio is larger than one, then the
  object is more likely to belong to class $i$ than to class $j$.

   Applied to the problem at hand, this can be written as,
\begin{equation}
\frac{P(c_i|\theta)}{P(c_j|\theta)} =
\frac{(N_i'/N_i)(N_i/N)}{(N_j'/N_j)(N_j/N)} =
\frac{N_i'}{N_j'}
\end{equation}
  where $N_i$ is the total number of objects in the training set belonging to
  class $i$, $N$ is the number of total objects in the training set, and a $'$
  denotes objects possessing observables $\theta$.  The $\theta$ represents a
  subset of all possible observables, which is equivalent to marginalizing
  over all other observables, and the set used differs according to which
  class of background is being tested.

   The criterion used by the autoscanner to identify objects of class $i$ is,
\begin{equation}
\frac{N_i'}{N_j'} >
\frac{N_i}{N_j}
\end{equation}
  where $i$ represents the classes of background (artifact, mover, dipole),
  and class $j$ is always taken as \sn.  The motivation for this criterion is
  that we are searching for an {\it overdensity} of objects of class $i$ in
  the region of observable space in question. In practice the threshold is a
  free parameter that can be adjusted to control the relative level of
  accepted signal to background events.

\begin{figure} [t]
   \begin{center}
   \includegraphics[angle=0, width=.7\textwidth]{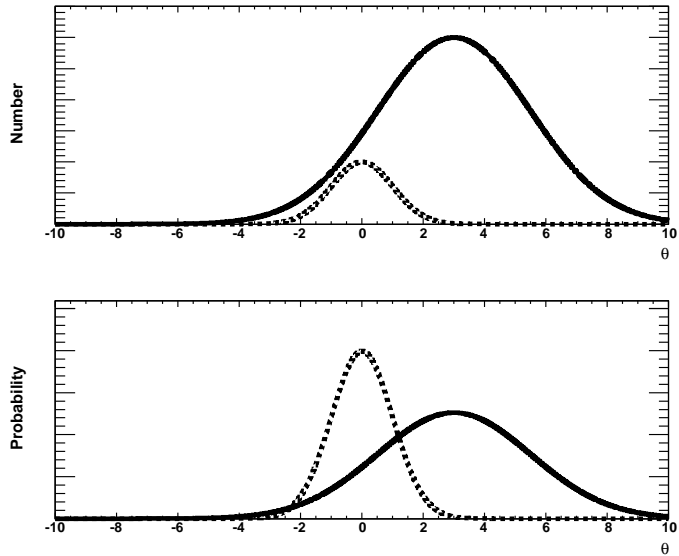}
   \end{center}
   \caption{Hypothetical distributions of an observable for signal (dashed)
            and background (solid). The peak of the signal occurs at 0;
            however, the vastly greater number of background events swamps the
            signal.  While it is true that an object with a value of this
            observable of 0 is more likely to be a background event, this is
            clearly not the behavior we want from our classifier, as we are
            trying to isolate the relatively rare signal events (i.e. \sn\
            epochs). The effect of recasting the problem from one of relative
            probabilities to one of overdensities.  The peak of the signal
            (dashed) is now strongly inconsistent with background.}

\label{fig:cartoon}
\end{figure}

  In this experiment, background events dominate over \sne\ epochs by a large
  factor, and a Bayesian classifier would reject \sn\ epochs that lie in a
  reasonable region of the observable space if the region overlaps with
  background events, even far away from the peak of the background
  distribution.  A cartoon of the situation is shown in
  Figure~\ref{fig:cartoon}.  Thus, the autoscanner can also be regarded as a
  Bayesian classifier where the thresholds are adjusted {\it a priori} to
  mitigate false negatives, at the cost of more background in the scanning.

  The training set for the autoscanner consists of all objects from the 2005
  observing season that were ranked by a human as an artifact, a dipole, a
  moving object, or a \sn.  This set is comprised of 91220 total objects, of
  which 65576 are a sub-class of background, and 25644 are objects ranked as a
  \sn.  The quantities used for classification of artifacts are, a measure of
  the objects ellipticity, the signal-to-noise ratio and the value of the
  reduced $\chi^{2}$ when fitting the object to a model of the \psf. In
  computing the $\chi^2$, we do not account for possible error in the \psf\
  model and so objects with large signal-to-noise values also have large
  $chi^2$ values.  Therefore, in classifying artifacts it is important to
  retain the correlation between $chi^2$ and signal-to-noise ratio.  The
  quantities used for classifying dipoles are the measured flux in negative
  valued pixels, and the ratio of measured flux in negative valued pixels to
  the flux in positive valued pixels.  The quantities used for classifying
  moving objects are the measured magnitudes, the magnitude of the apparent
  motion between different filters, and an angle describing the apparent
  motion between filters.  If the object has a detection in the 3 filter-bands
  $gri$, then the angle is the angle between the ``vector'' that describes the
  apparent motion between the $r$ and $i$ filters, and the ``vector'' that
  describes the apparent motion between the $i$ and $g$ filters.  If the
  object is truly moving then these apparent motions should be nearly
  co-linear.  If the object is only detected in 2 of the $gri$ filters then
  the angle is the angle between the apparent motion between the 2 filters and
  the ``unit vector'' aligned in the direction of increasing right-ascension.
  Main-belt asteroids typically have an apparent motion in a fixed direction,
  and using this angle in the classification is a way to encode that
  information.

\section{SN Ia Light Curve Models}
\label{app:snia_lc}

  This appendix describes the \snia\ light curve models used for photometric
  typing, redshift estimate, and selection of targets for spectroscopic
  observations.  To calculate synthetic light curves of Branch-normal \sneia\
  at various values of \dmB, the $B$-band magnitude change in 15 days from
  $B$-band maximum \citep{phillips93}, we adopt the results of
  \citet{phillips99}, who found relationships between the peak absolute $BVI$
  magnitudes of nearby \sneia\ as a function of \dmB.  These authors
  parameterize the peak magnitude relative to that of a reference \snia\ with
  $\dme = 1.1$ using a quadratic function of the form,
\begin{equation}
  \Delta M_{\rm{max}} = a \times [\Delta m_{15}(B) - 1.1] + b \times [\Delta
  m_{15}(B) - 1.1]^2.
\end{equation}

  There are $a$ and $b$ coefficients for each of the $BVI$ filters, as listed
  in Table~3 of \citet{phillips99}.  The $a$ coefficients ($a_B = 0.79$, $a_V
  = 0.67$, and $a_I = 0.42$) decrease as a function of the filter effective
  wavelength.  This shows that \sneia\ with larger \dmB\ are dimmer and
  redder; those with smaller \dmB\ are intrinsically more luminous and bluer.
  We find that the $a$ coefficient as a function of wavelength can be
  well-represented by the following linear relation,
\begin{equation}  
  \label{equ:acoeff}
  a(\lambda) = 1.248 - 1.045 \times 10^{-4}~\lambda(\mbox{\AA}).
\end{equation}
  We use this relation to perform color corrections directly onto the template
  spectra.  Operationally, defining $F_\nu(\lambda, x=0)$ to be the
  monochromatic flux (\fnu) of the standard template spectrum at rest-frame
  wavelength $\lambda$, then the luminosity-corrected value is given by,
\begin{equation}  
  \label{equ:fcorr}
  F_\nu(\lambda, x) \longrightarrow F_\nu(\lambda, x=0) \times 10^{-0.4~(ax +
    bx^2)},
\end{equation}
  where $x \equiv$ \dmB\ $- 1.1$, $b = 0.633$ \citep{phillips99}, and $a$ is
  given in Equation~\ref{equ:acoeff}.  Although this color correction is valid
  at $B$-band maximum, we assume that it is valid at all epochs.  This
  provides a way to compute synthetic light curves in the \ugriz\ system (or
  any other filter set) for a \snia\ with a given \dmB\ at arbitrary
  redshifts.

\begin{figure}[tb]
  \begin{center}
  \includegraphics[angle=270, width=.7\textwidth]{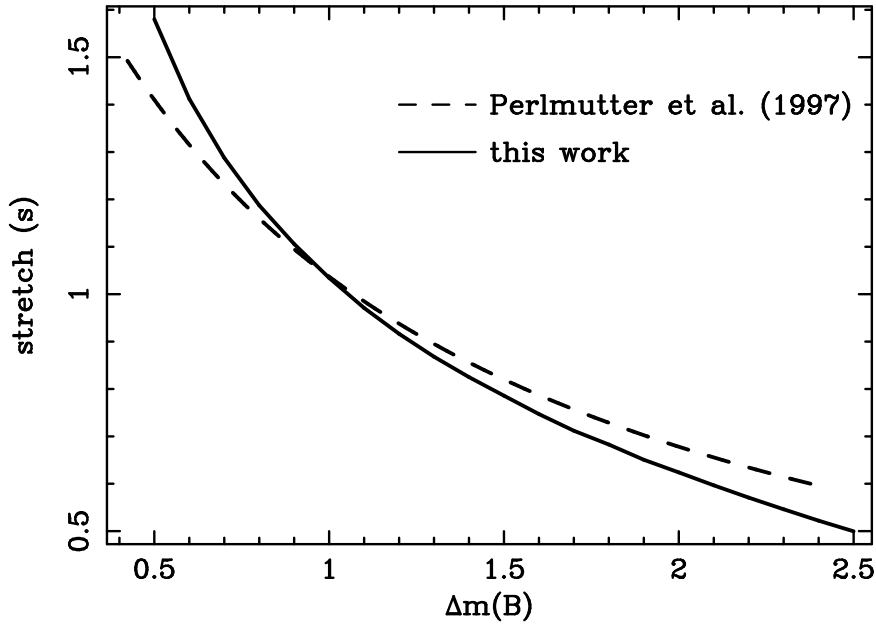}
  \end{center}
   \caption{The stretch factor $s$ at $z = 0$ as a function of \dmB\ adopted
     in this work (black) in comparison to that of \citet{perlmutter97}
     represented by a dashed line.} \label{fig:stretch}
\end{figure}

\begin{figure}[tb]  
  \begin{center}
    \includegraphics[angle=270, width=.7\textwidth]{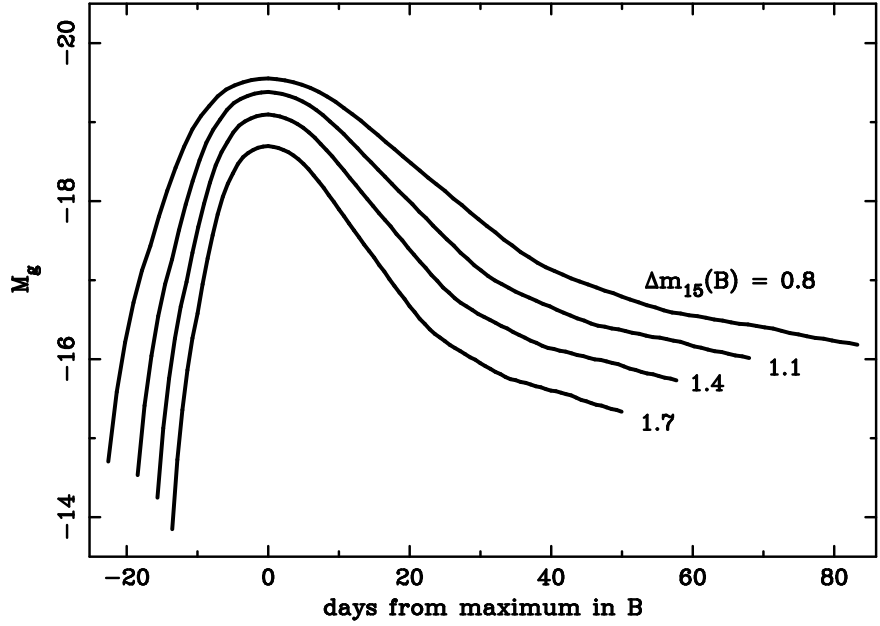}
  \end{center}
   \caption{Sample \snia\ rest-frame $g$-band light curves for four values of
     \dmB\ $= 0.8, 1.1, 1.4$, and 1.7.}
  \label{fig:model_lc}
\end{figure}

\begin{figure}[tb]  
  \begin{center}
    \includegraphics[angle=270, width=.7\textwidth]{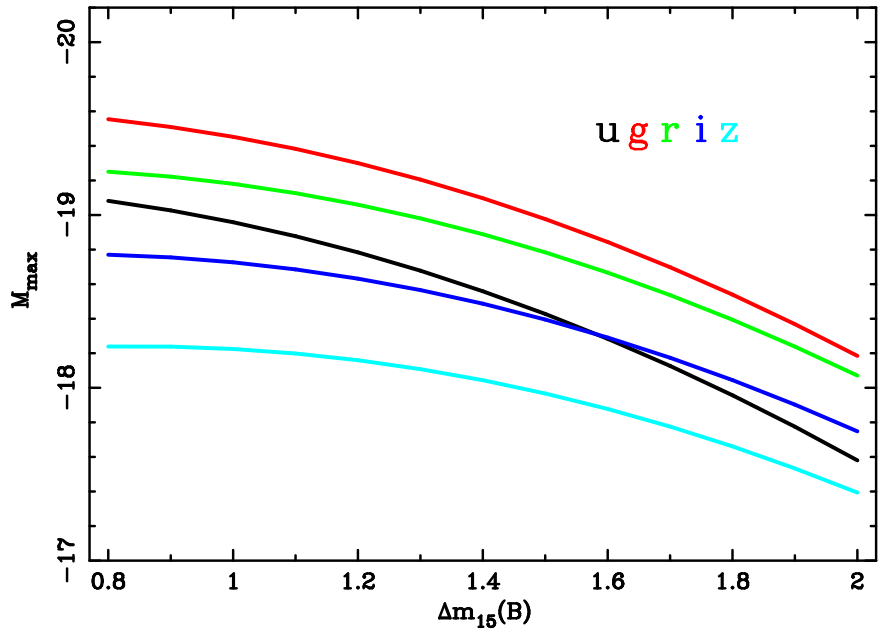}
  \end{center}
  \caption{Resulting peak absolute magnitudes for a \snia\ at zero redshift in
    the \sdss\ \ugriz\ system as functions of \dmB.  The curves are analogous
    to those shown in Figure~8 of \citet{phillips99}.}
  \label{fig:peakmag}
\end{figure}

  Finally, we stretch the light curves according to the method described in
  \citet{perlmutter97}.  The Nugent templates folded through the $B$-band
  filter transmission curve yield \dmB\ $= 1.05$, so we define $s$ to be unity
  when \dmB\ $= 1.05$.  For other values of \dmB, we express the stretch
  factor as $s = (1+z)~(15/\tau)$, where we have factored out the cosmological
  time delay factor, and $\tau$ is represented by the following 3rd-order
  polynomial,
\begin{equation}  
  \tau[\dme] = c_0 + c_1~\Delta m_{15}(B) + c_2~[\Delta m_{15}(B)]^2 +
  c_3~[\Delta m_{15}(B)]^3,
\end{equation}
  where $c_0 = 3.455$, $c_1 = 13.719$, $c_2 = -3.601$, and $c_3 = 0.946$.
  This gives an adequate representation of the $B$-band light curve for \dmB\
  between 0.5 and 2.5.  This function $s[$\dmB$]$ at zero redshift is plotted
  in Figure~\ref{fig:stretch}.  The same relation derived by
  \citet{perlmutter97} is also shown for comparison.
  Figure~\ref{fig:model_lc} shows rest-frame $g$-band light curves for
  different values of \dmB.  Peak absolute magnitudes in \ugriz\ as function
  of \dmB\ are shown in Figure~\ref{fig:peakmag}.

  Although this procedure does not always accurately reproduce the multi-band
  light curves, most notably the variation of the secondary peak in the red
  with \dmB\ \citep{riess96b}, they appear to be of sufficient quality for the
  purposes of photometric typing.

\clearpage

\begin{deluxetable}{lcrr}
  \tablewidth{0pt}
  \tablecaption{Stripe 82 footprint\label{tbl:stripe82}}
  \tablehead{
    \colhead{Strip} &
    \colhead{camcol} &
    \colhead{$\delta_{\rm{min}}$\tablenotemark{a}} &
    \colhead{$\delta_{\rm{max}}$\tablenotemark{a}}
  }
  \startdata

  82N     & 1 & $-1.058^\circ$ & $-0.833^\circ$ \\
  \nodata & 2 & $-0.638^\circ$ & $-0.413^\circ$ \\
  \nodata & 3 & $-0.219^\circ$ & $+0.007^\circ$ \\
  \nodata & 4 & $+0.201^\circ$ & $+0.426^\circ$ \\
  \nodata & 5 & $+0.621^\circ$ & $+0.846^\circ$ \\
  \nodata & 6 & $+1.041^\circ$ & $+1.266^\circ$ \\
  82S     & 1 & $-1.267^\circ$ & $-1.042^\circ$ \\
  \nodata & 2 & $-0.848^\circ$ & $-0.622^\circ$ \\
  \nodata & 3 & $-0.428^\circ$ & $-0.203^\circ$ \\
  \nodata & 4 & $-0.008^\circ$ & $+0.217^\circ$ \\
  \nodata & 5 & $+0.411^\circ$ & $+0.637^\circ$ \\
  \nodata & 6 & $+0.831^\circ$ & $+1.057^\circ$ \\

  \enddata

  \tablenotetext{a}{These values correspond to the nominal $\delta$ values.
    The exact telescope pointing differs slightly from night to night
    resulting in a shift in $\delta$ by as much as $0.003^\circ$ in either
    direction.}

\end{deluxetable}

\begin{deluxetable}{lcllrr}
  \tabletypesize{\scriptsize}
  \tablewidth{0pt}
  \tablecaption{2005 \sdss\ Supernova Imaging Runs\label{tbl:runs2005}}
  \tablehead{
    \colhead{Run} &
    \colhead{Strip} &
    \colhead{MJD$_{\rm{start}}$} &
    \colhead{MJD$_{\rm{end}}$} &
    \colhead{RA$_{\rm{start}}$} &
    \colhead{RA$_{\rm{end}}$}
  }
  \startdata
  5566 & N &  53616.215 &  53616.469 & $-32.72$ & $+58.91$ \\ 
  5582 & S &  53622.207 &  53622.477 & $-40.36$ & $+57.41$ \\ 
  5590 & N &  53623.156 &  53623.324 & $-50.67$ & $ +9.96$ \\ 
  5597 & S &  53625.176 &  53625.266 & $-50.62$ & $-18.43$ \\ 
  5603 & N &  53626.141 &  53626.449 & $-50.62$ & $+60.17$ \\ 
\smallskip
  5607 & S &  53627.137 &  53627.441 & $-50.59$ & $+59.90$ \\ 
  5610 & N &  53628.129 &  53628.434 & $-50.57$ & $+60.21$ \\ 
  5619 & S &  53634.129 &  53634.438 & $-50.68$ & $+60.71$ \\
\enddata

\end{deluxetable}

\begin{deluxetable}{lcllrr}
  \tabletypesize{\scriptsize}
  \tablewidth{0pt}
  \tablecaption{2006 \sdss\ Supernova Imaging Runs\label{tbl:runs2006}}
  \tablehead{
    \colhead{Run} &
    \colhead{Strip} &
    \colhead{MJD$_{\rm{start}}$} &
    \colhead{MJD$_{\rm{end}}$} &
    \colhead{RA$_{\rm{start}}$} &
    \colhead{RA$_{\rm{end}}$}
  }
  \startdata
  6281 & N &  53974.141 &  53974.203 & $-56.82$ & $-34.52$ \\ 
  6283 & N &  53974.270 &  53974.336 & $-11.69$ & $+12.42$ \\ 
  6287 & S &  53975.133 &  53975.465 & $-58.78$ & $+60.70$ \\ 
  6293 & N &  53977.371 &  53977.387 & $ +7.31$ & $+13.45$ \\ 
  6313 & N &  53989.121 &  53989.160 & $-60.05$ & $-45.99$ \\
  6314 & N &  53989.176 &  53989.473 & $-47.39$ & $+60.70$ \\ 
  6330 & S &  53990.312 &  53990.387 & $ +1.17$ & $+27.82$ \\ 
  6348 & S &  53993.277 &  53993.371 & $-31.19$ & $ +2.94$ \\
\enddata

\end{deluxetable}

\begin{deluxetable}{lr}
  \tablewidth{0pt}
  \tablecaption{Reduction of 2006 Handscanning by Autoscanner\label{tbl:scanning}}
  \tablehead{
    \colhead{} &
    \colhead{number of objects}
  }

  \startdata
  Total to be scanned\tablenotemark{a}   &     145,428 \\
  \hline
  1st epoch                              &     118,584 \\
  \hspace{0.5cm}Bright\tablenotemark{b}  &      21,882 \\
  \hspace{0.5cm}Scanned\tablenotemark{c} &       8,373 \\
  2nd epoch                              &      26,844 \\
  \hspace{0.5cm}Scanned                  &       6,007 \\
  \hline
  Total scanned\tablenotemark{d}         &      14,430 \\
  \enddata

  \tablenotetext{a}{The number of objects that humans would have scanned in
  2006 without the autoscanner.}
  \tablenotetext{b}{Objects that were detected for the first time and brighter
  than 21 mag in $g$ or $r$.}
  \tablenotetext{c}{Actual number of ``Bright'', first-epoch objects that were
  scanned.}
  \tablenotetext{d}{Total number of objects scanned in 2006.  This number
  differs slightly from the sum of the scanned first- and second-epoch objects
  because some scanners on a few occasions chose to scan first-epochs objects
  as faint as $21.5$~mag in $g$ or $r$.}

\end{deluxetable}

\begin{deluxetable}{ccrrrclc}
  \tabletypesize{\scriptsize}
  \tablewidth{0pt}
  \tablecaption{List of spectroscopically confirmed supernovae in
                2005\label{tbl:sn2005}}
  \tablehead{
    \colhead{Candidate} &
    \colhead{IAU} &
    &
    \colhead{RA\tablenotemark{c}} &
    \colhead{Dec} &
    \colhead{Discovery} &
    &
    \colhead{\# of} \\

    \colhead{ID\tablenotemark{a}} &
    \colhead{Name} &
    \colhead{z\tablenotemark{b}} &
    \multicolumn{2}{c}{(degrees in J2000)} &
    \colhead{Date} &
    \colhead{Type\tablenotemark{d}} &
    \colhead{Spectra\tablenotemark{e}}
  }
  \startdata
    722 &   2005ed &     0.0859 &      0.70569 &     +0.75127 & 2005-09-04 & SN Ia & 1 \\ 
    739 &   2005ef &     0.1072 &     14.59529 &     +0.67906 & 2005-09-04 & SN Ia & 1 \\ 
    744 &   2005ei &     0.1286 &  $-$30.80149 &     +0.31750 & 2005-09-04 & SN Ia & 1 \\ 
    762 &   2005eg &     0.1912 &     15.53535 &   $-$0.87899 & 2005-09-04 & SN Ia & 1 \\ 
    774 &   2005ex &     0.0936 &     25.46351 &   $-$0.87638 & 2005-09-04 & SN Ia & 2 \\ 
   1032 &   2005ez &     0.1299 &     46.79571 &     +1.11953 & 2005-09-06 & SN Ia & 1 \\ 
   1112 &   2005fg &     0.2585 &  $-$20.98238 &   $-$0.37519 & 2005-09-10 & SN Ia & 1 \\ 
   1114 &   2005lb &     0.0257 &  $-$16.29145 &   $-$0.25250 & 2005-09-10 & SN II & 1 \\
\enddata
  \tablenotetext{a}{Internal SN candidate designation.}
  \tablenotetext{b}{Final spectroscopic redshifts.  When narrow galaxy lines
  are detected in at least one of the spectra, the redshifts are given to four
  decimal places.  The typical uncertainties are $\sim 0.0005$.  When only the
  SN is detected, the redshifts are given to three decimal places and the
  uncertainties are $\sim 0.005$ (see \citealt{zheng07}).}
  \tablenotetext{c}{Right Ascension is provided in decimal degrees defined in
  the range [-180$^\circ$,+180$^\circ$].}
  \tablenotetext{d}{The final spectroscopic type of the SN candidate.
  Spectroscopically probable SNe Ia are designated ``SN Ia?'' (see
  \citealt{zheng07}).}
  \tablenotetext{e}{This includes all spectral observations performed
  including those of the SN candidate and possibly its host galaxy.}
  \tablenotetext{f}{Spectroscopically confirmed by \citet{quimby07}.}
  \tablenotetext{g}{Spectroscopically confirmed by the ESSENCE Group.}

\end{deluxetable}

\begin{deluxetable}{ccrrrclc}
  \tabletypesize{\scriptsize}
  \tablewidth{0pt}
  \tablecaption{List of spectroscopically confirmed supernovae in
                2006\label{tbl:sn2006}}
  \tablehead{
    \colhead{Candidate} &
    \colhead{IAU} &
    &
    \colhead{RA\tablenotemark{c}} &
    \colhead{Dec} &
    \colhead{Discovery} &
    &
    \colhead{\# of} \\

    \colhead{ID\tablenotemark{a}} &
    \colhead{Name} &
    \colhead{z\tablenotemark{b}} &
    \multicolumn{2}{c}{(degrees in J2000)} &
    \colhead{Date} &
    \colhead{Type\tablenotemark{d}} &
    \colhead{Spectra\tablenotemark{e}}
  }
  \startdata
  12778 &   2006fs &      0.10 &  $-$42.50417 &     +0.40879 & 2006-08-27 & SN Ia & 1 \\ 
  12779 &   2006fd &      0.08 &  $-$50.52813 &     +1.22112 & 2006-08-27 & SN Ia & 1 \\ 
  12780 &   2006eq\tablenotemark{f} &      0.05 &  $-$37.84531 &     +1.22811 & 2006-08-27 & SN Ia & 2 \\ 
  12781 &   2006er &      0.08 &      5.40656 &   $-$1.00997 & 2006-08-27 & SN Ia & 1 \\ 
  12782 &   2006fq &      0.07 &      5.00338 &   $-$0.62497 & 2006-08-27 & SN II & 1 \\ 
  12819 &   2006ho &      0.11 &      8.36686 &   $-$0.32024 & 2006-08-29 & SN II & 1 \\ 
  12820 &   2006fg &      0.04 &      9.58353 &   $-$0.29499 & 2006-08-29 & SN II & 4 \\ 
  12841 &   2006gk &      0.29 &  $-$38.65275 &   $-$1.03367 & 2006-09-11 & SN Ia & 1 \\
\enddata
  \tablenotetext{a}{Internal SN candidate designation.}
  \tablenotetext{b}{Preliminary spectroscopic redshifts.  The typical
  uncertainties are $\sim 0.01$. }
  \tablenotetext{c}{Right Ascension is provided in decimal degrees defined in
  the range [-180$^\circ$,+180$^\circ$].}
  \tablenotetext{d}{The preliminary spectroscopic type of the SN candidate.
  Spectroscopically probable SNe Ia are designated ``SN Ia?'' (see
  \citealt{zheng07}).}
  \tablenotetext{e}{This includes all spectral observations performed
  including those of the SN candidate and possibly its host galaxy.}
  \tablenotetext{f}{Spectroscopically confirmed by the CfA SN Group.}
  \tablenotetext{g}{Spectroscopically confirmed by the ESSENCE Group.}

\end{deluxetable}

\begin{deluxetable}{lrr}
  \tablewidth{0pt}
  \tablecaption{Summary of the 2005 and 2006 search seasons\label{tbl:summary}}
  \tablehead{
    \colhead{Number of} &
    \colhead{2005} &
    \colhead{2006} 
  }
  \startdata
  Nights on SDSS 2.5m             &      59 &       60 \\
  Runs                            &      73 &       90 \\
  Total number of objects         & 375,187 &  391,435 \\
  \hspace{0.5cm}Number of objects scanned       & 155,616 &   14,430 \\
  \hspace{0.5cm}Number of objects tagged as SN  &  24,402 &    4,189 \\
  \hline
  SN candidates discovered\tablenotemark{a}  &  11,385 &  3,694 \\
  \hspace{0.5cm}Single-epoch candidates\tablenotemark{b}      &   6,618 &    599 \\
  \enddata

  \tablenotetext{a}{The number of unique SN candidates tagged by a scanner.
  This number differs from the number of objects tagged as a SN because it
  includes multiple objects from the same candidate.}
  \tablenotetext{b}{Candidates that were detected in only a single run.  Most
  of these sources are likely to be solar system objects.}

\end{deluxetable}

\begin{deluxetable}{lrr}
  \tabletypesize{\scriptsize}
  \tablewidth{0pt}
  \tablecaption{Spectroscopic Observations of 2005 and 2006
                Candidates\label{tbl:spectro}}
  \tablehead{
    \colhead{Number of Spectroscopic Observations} &
    \colhead{2005} &
    \colhead{2006} 
  }
  \startdata
  SN Ia candidates\tablenotemark{a}                               &  173  &  241  \\
  \hspace{0.5cm}Identified as SN Ia              &  130\tablenotemark{b}  &  197\tablenotemark{c}  \\
  \hspace{0.5cm}Identified as probable SN Ia     &   16  &   14 \\
  \hspace{0.5cm}Resulting in host galaxy spectra of photometric SN Ia &    7  &    7 \\
  \hspace{0.5cm}Resulting in other host galaxy spectra &    3  &    6  \\
  \hspace{0.5cm}Identified as SN Ib/c                  &    1  &    0  \\
  \hspace{0.5cm}Identified as SN II                    &    5\tablenotemark{d}  &    6  \\
  \hspace{0.5cm}Identified as AGN                      &   2  &    0  \\
\enddata

  \tablenotetext{a}{Number of candidates targeted for spectroscopy that had
  been photometrically identified as a possible SN Ia.}
  \tablenotetext{b}{This includes one SN Ia that was spectroscopically
  confirmed by the Texas Supernova Search \citep{quimby07}.}
  \tablenotetext{c}{This includes five SNe Ia that were spectroscopically
  confirmed by the CfA SN Group and the ESSENCE Group.}
  \tablenotetext{d}{This includes one SN II that was spectroscopically
  confirmed by the ESSENCE Group.}
  \tablenotetext{e}{This includes observations of core-collapse SNe, novae,
  and other interesting transients.}
  \tablenotetext{f}{Of the 113 host galaxy spectra obtained, the measured
  redshifts of a sample of 81 sources were consistent with the photometric
  redshift estimated from the SN Ia model light curve fits, and were thus
  identified as photometric SN Ia.}
  \tablenotetext{g}{These were targeted as host galaxies of photometric SN Ia
  candidates.  The observations, however, resulted in either noisy,
  unclassifiable spectra or measured spectroscopic redshifts that are
  inconsistent with the SN Ia light curve redshifts implying that the
  candidates are not SNe Ia.}
  \tablenotetext{h}{The total number of duplicate and multi-epoch spectra; not
  the number of candidates targeted for multi-epoch spectroscopy.}

\end{deluxetable}

\end{document}